\documentclass[12pt]{article}
\usepackage[english]{babel}
\usepackage{indentfirst,graphicx,newlfont}
\usepackage{amssymb,amsmath,latexsym,amsthm,bm}
\usepackage{hyperref}
\usepackage{longtable}
\usepackage{url}
\usepackage{color}
\begin{document}
\thispagestyle{empty}
\newtheorem{df}{Definition}
\newtheorem{prop}{Proposition}
\newtheorem{remark}{Remark}
\newtheorem{cor}{Corollary}
\newtheorem{es}{Example}
\newtheorem{teo}{Theorem}
\newtheorem{lm}{Lemma}
\newtheorem{pr}{Property}
\bmdefine{\bi}{i}
\bmdefine{\bj}{j}
\def\T{{\footnotesize {^{_{\sf T}}}}} 
\newcommand{\Real}{{\rm I}\negthinspace {\rm R}}
\newcommand{\sthat}{\widehat}
\newcommand{\I}{\mathbb{I}}

\newtheorem{theorem}{Theorem}[section]
\newtheorem{lemma}[theorem]{Lemma}
\newtheorem{proposition}[theorem]{Proposition}
\newtheorem{corollary}[theorem]{Corollary}

\author{F.\ Giummol\`e$^+$ and V.\ Mameli$^+$\\
{\it {\small $^+$Dept.\ of Environmental Sciences, Informatics and Statistics, Ca' Foscari University of Venice, Italy}}
\\[-.7ex]
{\tt {\small giummole@unive.it,valentina.mameli@unive.it}} \\
E. \ Ruli$^*$ and L.\ Ventura$^*$\\
 {\it {\small $^*$Dept. of Statistical Sciences, University of Padova, Italy}} \\
 {\tt {\small ruli@stat.unipd.it, ventura@stat.unipd.it}}}
 
\title{Objective Bayesian inference with proper scoring rules}
\maketitle
\begin{abstract}
Standard Bayesian analyses can be difficult to perform when the full likelihood, and consequently the full posterior distribution, is too complex or even impossible to specify or if robustness with respect to data or to model misspecifications is required. In these situations, we suggest to resort to a posterior distribution for the parameter of interest based on proper scoring rules. Scoring rules are loss functions designed to measure the quality of a probability distribution for a random variable, given its observed value. Important examples are the Tsallis score and the Hyv\"arinen score, which allow us to deal with model misspecifications or with complex models. Also the full and the composite likelihoods are both special instances of scoring rules. 

The aim of this paper is twofold. Firstly, we discuss the use of scoring rules in the Bayes formula in order to compute a posterior distribution, named SR-posterior distribution, and we derive its asymptotic normality. Secondly,  we propose a  procedure for building default priors for the unknown parameter of interest, that can be used to update the information provided by the scoring rule in the SR-posterior distribution. In particular, a reference prior is obtained by maximizing the average $\alpha-$divergence from the SR-posterior distribution. For $0 \leq |\alpha|<1$, the result is a Jeffreys-type prior that is proportional to the square root of the determinant of the Godambe information matrix  associated to the scoring rule. Some examples are discussed.
\end{abstract}



\section{Introduction}

In the Bayesian setting, when the full likelihood is too complex and difficult to specify or when robustness with respect to data or to model misspecifications is required, several authors have proposed the use of surrogate likelihoods in the Bayes formula, in place of the full likelihood.  Although this approach is non-orthodox, it is widely used in the recent statistical literature and theoretically motivated in several papers. See, among others, Pauli {\em et al.} (2011) and Ribatet {\em et al.} (2012) for the use of composite likelihoods, Lazar (2003), Lin (2006), Schennach (2005), Greco {\em et al.} (2008), Chang and Mukerjee (2008), Ventura {\em et al.} (2010), and Yang and He (2012) for the use of empirical and quasi-likelihoods; see also the review by Ventura and Racugno (2016), which addresses the use of pseudo-likelihoods in the Bayesian framework.

To deal with complex models or model misspecifications, useful surrogate likelihoods can be obtained trough proper scoring rules. A scoring rule (see, for instance, the recent overviews by Machete, 2013, and Dawid and Musio, 2014, and references therein) is a special kind of loss function designed to measure the quality of a probability distribution for a random variable, given its observed value. It is proper if it encourages honesty in the probability evaluation. Proper scoring rules supply unbiased estimating equations for any statistical model, which can be chosen to increase robustness or for ease of computation. The Brier score (Brier, 1950), the logarithmic score (Good, 1952), the Tsallis score (Tsallis, 1988), and the Hyv\"arinen score (Hyv\"arinen, 2005) are widely known. In particular, when using the logarithmic score, the full likelihood and the composite likelihood (Varin {\em et al.}, 2011) are obtained as special cases of proper scoring rules (see for instance Dawid and Musio, 2014). While frequentist scoring rule inference has been widely discussed (see Dawid {\em et al.}, 2016, and references therein), Bayesian inference is relatively unexplored. Few exceptions are Dawid and Musio (2015) who focus on  Bayesian model selection, Ghosh and Basu (2016) who consider robust Bayes estimation using the density power divergence measure, and Musio {\em et al.} (2017) who give an illustration of Bayesian inference for directional data through the Hyv\"arinen score; see also Pauli {\em et al.} (2011) and Ribatet {\em et al.} (2012) for the use of composite likelihoods in the Bayes formula.

To perform Bayesian inference, a suitable prior distribution on the parameter of interest must be elicited. In this paper we focus on default priors which are frequently used in Bayesian applications and which are still an active area of research (see, among others, Berger, 2006, Berger {\em et al.}, 2009, 2012, Ghosh, 2011, Walker, 2016, Leisen {\em et al.}, 2017). The two most common objective priors are the Jeffreys prior (Jeffreys, 1961), which uses the information about the parameter contained in the Fisher information, and the reference prior (Bernardo, 1979), which is based on the maximization of a distance in information between the prior and the posterior.

The goal of this paper is twofold. The first aim is to discuss the use of scoring rules in order to compute a posterior distribution, called here the SR-posterior distribution. In particular, we suggest a SR-posterior distribution obtained by extending to the general scoring rule setting the curvature adjustment of the composite likelihood proposed by Chandler and Bate (2007) and Ribatet {\em et al.} (2012). The calibration is necessary in order to show that the SR-posterior distribution is, up to order $O_p(n^{-1/2})$, normally distributed, with the same asymptotic variance of the scoring rule estimator. Indeed, a calibrated scoring rule is needed to reach the right asymptotic variance in the normal approximation, as well as a correct shape of the posterior distribution. 

The second aim is to propose the elicitation of a default prior for the unknown parameter of interest. In particular, we focus on reference priors as pioneered by Bernardo (1979); for a review see Bernardo (2005) and Ghosh (2011). Our purpose is to construct reference priors obtained by maximizing $\alpha-$divergences  from the SR-posterior distribution. The $\alpha-$divergences are a well known class of discrepancy functions which include as a special case the Kullback-Leibler divergence. We show that, for $0 \leq |\alpha| <1$,  the maximizer is a Jeffreys-type prior that is proportional to the square root of the determinant of the Godambe information matrix.

The paper unfolds as follows. Section 2 gives background on scoring rules. Section 3 discusses the properties of the SR-posterior distribution. Section 4 presents the construction of reference priors based on $\alpha-$divergences using SR-posteriors. Examples are presented in Section 5 both in the context of complex models and in the context of robustness theory. 
Further suggestions and comments can be found in the  conclusions.


\section{Scoring rules}

A scoring rule is a loss function designed to measure the quality of a given probability distribution $Q$ for a random variable $X$, in view of the result $x$ of $X$; see Dawid (1986). The function $S(x;Q)$ takes values in $\mathbb{R}$ and its expected value under $P$ will be denoted by $S(P;Q)$. The scoring rule $S$ is called proper relative to the class of distributions $\mathcal{P}$ if  the following inequality is satisfied for all $P,\,Q\in\mathcal{P}$:
\begin{equation}
\label{eq:1}
S(P;Q)\geq S(P;P).
\end{equation}
It is strictly proper relative to $\mathcal{P}$ if equation (\ref{eq:1}) is satisfied with equality if and only if $Q = P$. Note that in the following we identify a distribution $Q$ by its probability density $q$ with respect some measure $\mu$; so the two notations $S(x;q)$ and $S(x;Q)$ are indistinguishable. 

An important example of proper scoring rules is the logarithmic score, which is defined as $S(x;Q)=-\log{q(x)}$ (Good, 1952) and corresponds to minus the log-likelihood function. Other widely used proper scoring rules are the Tsallis and the Hyv\"arinen scores. The Tsallis score (Tsallis, 1988, Basu {\em et al.}, 1998) is given by
\begin{equation}
  \label{eq:tsallisscore}
  S(x;Q) = (\gamma - 1) \int\!  q(y)^\gamma \, d\mu(y) - \gamma q(x)^{\gamma-1},
  \quad \gamma>1 .
\end{equation}
The Tsallis score gives in general robust procedures (Basu {\em et al.}, 1998, Dawid {\em et al.}, 2016), and $\gamma$ is a tuning constant that governs the trade-off between efficiency and robustness. The Hyv\"arinen score in its original formulation (Hyv\"arinen, 2005) with $x\in\mathbb{R}^{d}$ is defined as
\begin{equation}
\label{hiv}
S(x,Q) = \varDelta \log{ q(x)} + \frac{1}{2}|| \nabla \log{ q(x)}||^2,
\end{equation}
where $||\cdot||$ is the standard norm, $\nabla$ denotes the gradient vector and $\varDelta$ the Laplacian operator. 


\subsection{Inference based on scoring rules} 

Let $x = (x_1,\ldots , x_n)$ be the realization of a random sample of size $n$ from a distribution with density function $p(x|\theta)$, with $\theta \in \Theta \subseteq \mathbb{R}^{d}$, ${d} \geq 1$. Moreover, let $L(\theta) = L(\theta;x)$ and $\ell(\theta)=\log L(\theta;x)$ be the likelihood and the log-likelihood functions, respectively. 

The validity of inference about $\theta$ using scoring rules can be justified by invoking the general theory of unbiased $M$-estimating functions. Indeed, inference based on proper scoring rules is a special kind of $M$-estimation (see, e.g., Dawid {\em et al.}, 2016, and references therein). The class of $M$-estimators is broad and includes a variety of well-known estimators. For example  it includes the maximum likelihood estimator (MLE), the maximum composite likelihood estimator (see e.g.\ Varin {\em et al.}, 2011), and robust estimators (see e.g.\ Hampel {\em et al.}, 1986, Huber and Ronchetti, 2009), among many others.

Given a proper scoring rule $S(x;P_\theta)=S(x;\theta)$, let us denote by $S(\theta)=\sum_{i=1}^n S(x_i;\theta)$ the total empirical score. Moreover,  let $s(x;\theta)$ be the gradient vector of $S(x;\theta)$ with respect to $\theta$, i.e.\ $s(x;\theta)=\partial S(x;\theta)/\partial \theta$. Under broad regularity conditions (see Mameli and Ventura, 2015, and references therein), the scoring rule estimator $\tilde\theta$ is the solution of the unbiased estimating equation 
$$
S_{\theta}(\theta) = \sum_{i=1}^n s(x_i;\theta) = 0\,,
$$ 
(see Dawid and Lautitzen, 2005, Dawid, 2007, Dawid {\em et al.}, 2016) and it is asymptotically normal, with mean $\theta$ and covariance matrix $V(\theta)/n$, with
\begin{eqnarray}
V (\theta) = K(\theta)^{-1} J(\theta) (K(\theta)^{-1}), 
\label{var}
\end{eqnarray}
where $K(\theta) = E_\theta (\partial s(X;\theta)/\partial \theta^{T})$ and $J(\theta) = E_\theta(s(X;\theta) s(X;\theta)^{T} )$ are the sensitivity and the variability matrices, respectively.  The matrix $G(\theta)=V(\theta)^{-1}$ is known as the Godambe information matrix (see Godambe, 1960) and its form is due to the failure of the second Bartlett identity (also called information identity) since, in general, $K(\theta) \neq J(\theta)$. In the special case of the logarithmic score, we have that $G(\theta)=K(\theta)=J(\theta)$ is the Fisher information matrix.

Asymptotic inference for scoring rule is usually based on first-order approximations, but several examples have shown the inaccuracy of these methods; see for instance Dawid et al. (2016) and Mameli and Ventura (2015). In order to improve the accuracy of first order methods, refinements have been obtained by resorting to higher-order asymptotic theory, which encompass the classical results for likelihood quantities while allowing for the failing of the information identity; see Mameli and Ventura (2015) and Mameli {\em et al.}\ (2017).


\section{Posterior distributions based on scoring rules} 

The use of surrogate likelihoods in the Bayes formula has received great attention in the last decade (see the review by Ventura and Racugno, 2016, and references therein). In particular, the use of the composite likelihood, in the Bayes formula has been considered for instance by Smith and Stephenson (2009), Pauli {\em et al.}\ (2011) and Ribatet {\em et al.}\ (2012). Since the composite likelihood does not satisfy the information identity, it is necessary to calibrate it in order to reach the right asymptotic variance in the normal approximation, as well as a correct shape of the posterior distribution. The correct curvature of the posterior distribution based on composite likelihoods can also be reached by using the composite score function in the Approximate Bayesian Computation (ABC) procedure, as discussed in Ruli {\em et al.}\ (2016).

While the use of surrogate likelihoods for Bayesian inference has been widely discussed in the recent literature to deal with complex models or robustness, the use of scoring rules in the Bayesian framework has not been deeply explored; exceptions are Dawid and Musio (2015), Ghosh and Basu (2016) and Musio {\em et al.}\ (2017).

Paralleling the derivation of posterior distributions based on composite likelihoods, a SR-posterior distribution can be obtained by using a scoring rule instead of the full likelihood in Bayes formula. However, since the scoring rule does not satisfy the information identity, it must be suitably calibrated before deriving the SR-posterior distribution.  In particular, here we suggest a SR-posterior distribution obtained by extending to the scoring rule setting the curvature adjustment of  the composite likelihood proposed by Chandler and Bate (2007) and Ribatet {\em et al.}\ (2012). In the following we study the asymptotic normality of the SR-posterior distribution; this result validates the use of scoring rule in a Bayes formula (see Ventura and Racugno, 2016, Section 3.3).

Let $\pi(\theta)$ be a prior distribution for the parameter $\theta$. The proposed  SR-posterior distribution  is defined as 
\begin{equation}
\label{star}
\pi_{SR}(\theta|x) \propto \pi(\theta) \exp{\{-S (\theta^*)\}},
\end{equation}
with $\theta^*= \theta^*(\theta)=\tilde{\theta}+C(\theta-\tilde{\theta})$, where $C$ is a $d\times d$ fixed matrix such that $C^T K(\theta)C=G(\theta)$. A possible choice of the matrix $C$ is given by $C = M^{-1}M_A$, with $M_A^{T}M_A = G$ and $M^{T}M = K$; for details, see Ribatet {\em et al.}\ (2012) and references therein. Note that in formula \eqref{star}, when $S(\theta)=-\sum_{i=1}^n\log p(x_i|\theta)$ is the logarithmic scoring rule, the posterior $\pi_{SR}(\theta|x)$ reduces to the classical posterior distribution based on the likelihood function, i.e.\ $\pi(\theta|x)\propto \pi(\theta)\exp{\{\sum_{i=1}^n\log p(x_i|\theta)\}}$. 

The choice of a prior distribution $\pi(\theta)$ to be used with a scoring rule involves the same problems typical of the standard Bayesian perspective, namely the elicitation of a proper prior distribution and the fact that default prior distributions can often be improper. The study of the possible impropriety of the SR-posterior has to be considered case-by-case, similarly to what happens for the genuine posterior distribution based on a default prior.

Under the same regularity conditions necessary for the asymptotic results for scoring rule inference (see for instance Mameli and Ventura, 2015) as $n \rightarrow \infty$, it can be shown that the scoring rule posterior distribution (\ref{star}) is, up to order $O_p(n^{-1/2})$, normally distributed with mean $\tilde{\theta}$ and variance $H(\tilde{\theta})^{-1}/n$, i.e.\
\begin{equation}
\label{posterior}
\pi_{SR}(\theta|x) \stackrel{\cdot}{\sim} N_{d} \left( \tilde{\theta},\frac{H(\tilde{\theta})^{-1}}{n} \right),
\end{equation} 
where $H(\theta)=C^{T}(\partial^2 S(\theta)/\partial\theta^2)C/n$. Note that $H(\tilde{\theta})$ converges almost surely to $G(\theta)$ as $n \rightarrow \infty$. This result is stated in the next Theorem \ref{uno} that gives the expansion of the scoring rule posterior distribution (\ref{star}) up to third order. 

Before stating Theorem \ref{uno}, we introduce the notation which will be used in the rest of the paper.  
We use indices to denote derivatives of $S(x;\theta)$ and $\pi(\theta)$ with respect to the components of the parameter $\theta$. Therefore, for example, $S_{jkh}(\theta)=\partial^3 S(\theta)/\partial \theta^j\partial\theta^k\partial\theta^h$, $S_{ijkh}(\theta)=\partial^4 S(\theta)/\partial \theta^i\partial \theta^j\partial\theta^k\partial\theta^h$, for $1\leq i,j,k,h\leq d$, $\pi_{i}(\theta)=\partial \pi(\theta)/\partial \theta^i$ and 
$\pi_{ij}(\theta)=\partial^2 \pi(\theta)/\partial \theta^i\partial\theta^j$. In the sequel, a tilde over a quantity means evaluation of that quantity in $\tilde\theta$. For example, $\tilde{\pi}=\pi(\tilde\theta)$ and $\tilde{H}=H(\tilde\theta)$. Further, $\tilde{S}_{jkh}=S_{jkh}(\tilde{\theta})$, $\tilde{S}_{ijkh}=S_{ijkh}(\tilde{\theta})$, $1\leq i,j,k,h\leq d$, $\tilde\pi_i=\pi_{i}(\tilde\theta)$ and $\tilde\pi_{ij}=\pi_{ij}(\tilde\theta)$. Let $c_{ij}$ and $\tilde{h}^{ij}$ be the elements of the matrices $C$ and $\tilde{H}^{-1}=H(\tilde\theta)^{-1}$, respectively. Moreover, we use the Einstein summation convention so that when an index appears twice in an expression, summation on that index is intended.

\begin{theorem}
\label{uno}
Let $w=(w^1,\dots,w^{d})^T=n^{1/2}(\theta-\tilde\theta)$. The SR-posterior distribution for $w$ can be written as
\begin{eqnarray}
\label{post-expansion}
\pi_{SR} (w|x)=\phi_d(w;\tilde H^{-1})
\left[
1+n^{-1/2}A_1(w) + n^{-1}A_2(w)\right]+O_p(n^{-3/2}),
\end{eqnarray}
where $\phi_d(w;\tilde H^{-1})$ is the density of a $d-$variate normal distribution with zero mean vector and variance matrix $
\tilde H^{-1}$, 
$$
A_1(w)= \frac{\tilde\pi_i}{\tilde\pi}w^i-\frac{1}{6}\frac{\tilde S_{ijk}}{n}c_{ir}c_{js}c_{kt}w^rw^sw^t
$$ 
and 
\begin{eqnarray*}
A_2(w)&=&
\frac{1}{2}\frac{\tilde\pi_{ij}}{\tilde\pi}(w^iw^j-\tilde h^{ij})\\
&&-\frac{1}{6}\frac{\tilde\pi_i}{\tilde\pi} \frac{\tilde S_{jkh}}{n}c_{jr}c_{ks}c_{ht}
(w^i w^r w^s w^t-3 \tilde h^{ir}\tilde h^{st})\\
&&-\frac{1}{24}\frac{\tilde S_{ijkh}}{n} c_{ir}c_{js}c_{kt}c_{hu}
(w^r w^s w^t w^u -3\tilde h^{rs}\tilde h^{tu})\\
&&+\frac{1}{72}\frac{\tilde S_{ijk}\tilde S_{rst}}{n^2}
\left[c_{ia}c_{jb}c_{kc}c_{rd}c_{se}c_{tf}(w^aw^bw^cw^dw^ew^f-9\tilde h^{ab}\tilde h^{cd}\tilde h^{ef})
\right.\\
&&\quad\quad\quad
\left.+c_{ia}c_{rb}c_{jc}c_{sd}c_{ke}c_{tf}(w^aw^bw^cw^dw^ew^f-6\tilde h^{ab}\tilde h^{cd}\tilde h^{ef})\right].
\end{eqnarray*}
\end{theorem}

The proof is given in the Appendix. 

In practice, Theorem \ref{uno} shows that the SR-posterior distribution (\ref{star}) has the correct curvature, namely the variance of the posterior distribution (\ref{posterior}) is asymptotically equivalent to that of the minimum scoring rule estimator. When in particular $S(x;\theta)$ is the logarithmic score, it can be shown that (\ref{post-expansion}) reduces to the expansion given in Datta and Mukerjee (2004) for the classic posterior distribution $\pi(\theta|x) \propto \pi(\theta) L(\theta;x)$.


\section{Reference priors obtained from $\alpha-$divergences}

The information on $\pi(\theta|x)$ induced by $\pi(\theta)$ may be measured in terms of a divergence $D(\cdot,\cdot)$ between the prior and the posterior distribution: the  higher the divergence, the lower the influence of the prior on the posterior. Let $D_{\pi}(x)=D(\pi(\theta),\pi(\theta|x))$, and let 
$p(x)$ be the marginal distribution of $X$.  
Minimizing the information in a prior is equivalent to maximize the expected divergence $D_\pi$ from the corresponding posterior, i.e.\ the functional 
\begin{equation}\label{functional}
T(\pi)=\int_{\mathcal{X}}D_\pi (x)p(x)dx.
\end{equation} 
Here, we focus on the well-known family of $\alpha-$divergences, defined as 
\[
D_{\pi}(x)=\frac{1}{\alpha(1-\alpha)}\int_{\Theta}\left\lbrace 1-\left(\frac{\pi(\theta)}{\pi(\theta|x)}\right)^{\alpha}\right\rbrace \pi(\theta|x)d\theta,
\]
which for $\alpha\rightarrow 0$ reduces to the Kullback-Liebler divergence, for $\alpha=1/2$ corresponds to twice the Hellinger distance and for $\alpha=-1$ is equivalent to the $\chi^2-$divergence. 

In this section, we extend the results of Ghosh {\em et al.} (2011) and Liu {\em et al.} (2014) to the context of scoring rules. In particular, in the following theorem we propose reference priors obtained by maximizing the expected $\alpha-$divergence from the SR-posterior distribution (\ref{star}).

In the sequel, we denote by $g_{ij}$ and $g^{ij}$ the components of matrices $G(\theta)$ and $G(\theta)^{-1}$ respectively. Moreover, let $E_\theta(S_{ijk}/n)=B^S_{ijk}(\theta)+o(n^{-1/2})$ and let $a^S=(a^S_1,\dots,a^S_{d})^T$, with $a^S_i=B^S_{khl}c_{ki}c_{hu}c_{lt}g^{ut}$. Similarly, we define $E_\theta(\ell_{ijk}/n)=B^\ell_{ijk}(\theta)+o(n^{-1/2})$ and $a^\ell=(a^\ell_1,\dots,a^\ell_{d})^T$, with $a^\ell_i=B^\ell_{ihk}i^{hk}$. Here, $\bi^{ij}$ are the components of the inverse of the Fisher information matrix $I=E_\theta(-\partial^2\ell(\theta)/\partial\theta\partial\theta^T)$. Finally, 
 $\Sigma=(\sigma^{ij})_{ij}$, with $\sigma^{ij}=nCov_\theta(\tilde\theta^i,\hat\theta^j)$, where $\hat\theta$ is the maximum likelihood estimator for $\theta$. Here, all indices range between $1$ and $d$. 

\begin{theorem}
\label{due}
When $0\leq |\alpha|<1$, the prior which asymptotically maximizes the expected $\alpha-$divergence to the SR-posterior distribution (\ref{star}) is  
\begin{eqnarray}\label{nostraprior}
\pi_G (\theta)\propto |G(\theta)|^{1/2}.
\end{eqnarray}
When $\alpha=-1$, the prior which asymptotically maximizes the expected $\chi^2-$divergence to the SR-posterior distribution (\ref{star}) is such that
\begin{equation}
\label{log-solution}
\frac{\partial\log\pi_{\chi}(\theta)}{\partial\theta}=
\frac{1}{4}\left[
6a^S G^{-1} + 4a^\ell I^{-1}+ |G|^{-1}\frac{\partial |G|}{\partial\theta}(5G^{-1}-4\Sigma)
+2G \nabla\cdot (5G^{-1}-4\Sigma) \right] \Gamma,
\end{equation}
where   $\nabla\cdot G^{-1}=
(\partial g^{1j}/\partial\theta^j, \dots, \partial g^{{d} j}/\partial\theta^j)^T$, $\nabla\cdot \Sigma=
(\partial \sigma^{1j}/\partial\theta^j, \dots, \partial \sigma^{{d} j}/\partial\theta^j)^T$ and $\Gamma=(G^{-1}-4I^{-1}+4\Sigma)^{-1}$.
\end{theorem}

The proof is given in the Appendix. 

It is important noticing that, if we use the full likelihood in the Bayes formula,  (\ref{nostraprior}) and (\ref{log-solution}) reduce to the results obtained in Liu {\em et al.} (2014) and, for the case  of a scalar parameter, in Ghosh {\em et al.} (2011).

As shown in the Appendix, for $|\alpha|>1$ and $\alpha=1$ a maximizer for the expected $\alpha-$divergence does not exist.

Note that, when dealing with the multidimensional parameter case, the use of the square root of the determinant of the Godambe information matrix may be questionable as in the case of the Jeffreys prior.  Following the advices for default priors, one possibility is to assume that the components of $\theta$ are a priori independent and to use the one-dimensional reference prior for each of the parameters. An alternative is to consider a sequential scheme as in Berger and Bernardo (1992).  On the other hand, it should be noted that the ordering of the parameters is relevant. Unless the practitioner has a specific ordering in mind, different orderings may lead to different reference priors.

For $\alpha=-1$, the preceding expression (\ref{log-solution}) depends on quantities related to both the scoring rule and the likelihood.  In practice this makes the proposed prior useful only when the likelihood function is available, as for instance in the context of robust inference. We also note that the required calculations remarkably simplify in the scalar parametric case.

The proposed prior distribution (\ref{nostraprior}) shares some important properties with the Jeffreys prior. The most relevant is invariance with respect to one-to-one changes in the parametrization. 
As shown in Liu {\em et al.} (2014), when the functional (\ref{functional}) is invariant, the maximizer is also invariant. Thus, invariance of the proposed prior distributions follows from the well known invariant properties of $\alpha$-divergences. 
Anyway, for $0\leq|\alpha|<1$, it can be easily seen that $G(\theta)$ is a second order tensor so that, if $\psi(\theta)$ is a new parametrisation, $G_{rs}(\psi)=G_{ij}(\theta(\psi))\theta^i_r(\psi) \theta^j_s(\psi)$, where $\theta^i_r=\partial\theta^i/\partial\psi^r$. Thus it follows that $\pi_G(\psi)=\pi_G(\theta(\psi))|\partial\theta(\psi)/\partial\psi|$.

We would like to stress that this is a general property that holds for every model and every scoring rule.
Other properties of the proposed prior distribution (\ref{nostraprior}) in general depend on the scoring rule under consideration. In the following we present two special examples.

\subsection{The Tsallis scoring rule}
 
Let $X$ be distributed as a location model with parameter $\mu\in\mathbb{R}$, so that $p(x|\mu)=p_0(x-\mu)$. The Tsallis scoring rule in this case takes the following form
\begin{eqnarray*}
S(x;\mu)&=&(\gamma-1)\int p_0(x-\mu)^\gamma dx-\gamma p_0(x-\mu)^{\gamma-1}\\
&=& \mbox{const}-\gamma p_0(x-\mu)^{\gamma-1}.
\end{eqnarray*}
Its derivatives with respect to $\mu$ are all functions of $x-\mu$, so that the corresponding expected values, i.e.  $K(\mu)$, $J(\mu)$ and  $G(\mu)$ are independent of $\mu$. Thus, $\pi_G(\mu)\propto 1$, as the Jeffreys prior.

Also for scale models, $\pi_G(\sigma)$ with the Tsallis score coincides with the Jeffreys prior. 
Indeed, let  $X$ be distributed as a scale model with parameter $\sigma>0$, so that $p(x|\sigma)=p_0(x/\sigma)/\sigma$. The Tsallis scoring rule in this case reduces to
\begin{eqnarray*}
S(x;\sigma)&=&(\gamma-1)\int \frac{1}{\sigma^\gamma} p_0\left(\frac{x}{\sigma}\right)^\gamma dx-\gamma \frac{1}{\sigma^{\gamma-1}} p_0\left(\frac{x}{\sigma}\right)^{\gamma-1}\\
&=& \frac{1}{\sigma^{\gamma-1}}\left[ \mbox{const}-\gamma p_0\left(\frac{x}{\sigma}\right)^{\gamma-1}\right].
\end{eqnarray*}
It is easy to show that the first three derivatives of $S(x;\sigma)$ can be written as
\[
S_\sigma=\frac{1}{\sigma^\gamma} f_1\left(\frac{x}{\sigma} \right), \quad
S_{\sigma\sigma}=\frac{1}{\sigma^{\gamma+1}} f_2\left(\frac{x}{\sigma} \right), \quad
S_{\sigma\sigma\sigma}=\frac{1}{\sigma^{\gamma+2}} f_3\left(\frac{x}{\sigma} \right),
\]
where $f_1$, $f_2$ and $f_3$ are suitable functions. Notice that the expectation of any function of $X/\sigma$ with respect to $p(x;\sigma)$ does not depend on $\sigma$ itself.  Thus, 
$J(\sigma)\propto 1/\sigma^{2\gamma}$ and $K(\sigma)\propto 1/\sigma^{\gamma+1}$, so that  $G(\sigma) \propto 1/\sigma^{2}$ and  $\pi_G(\sigma)\propto 1/\sigma$.

\subsection{Hyv\"arinen scoring rule} 

Let $X$ be a non-degenerate random variable belonging to the one-parameter natural exponential family
\begin{eqnarray}
p(x|\theta)&=&\exp{\{\theta x-k(\theta)+a(x)\}},\; \quad x\in\mathbb{R}
\ .
\label{expfam}
\end{eqnarray}
Let denote by $a'(x_i)=\partial a(x_i)/ \partial x_i$ and  $a''(x_i)=\partial^2 a(x_i)/ \partial x_i^2$. The Hyv\"arinen total empirical score in this case reduces to 
\begin{equation}\label{hyvexp}
 S(\theta)=-\left\lbrace 2\sum_{i=1}^na''(x_i)+\sum_{i=1}^n\left[  \theta +a'(x_i)\right] ^2\right\rbrace.
 \end{equation} The Hyv\"arinen score estimator for this family is
\begin{equation*}
\tilde\theta=-\frac{\sum_{i=1}^na'(X_i)}{n}
\ ,
\end{equation*}
which can be computed without knowledge of the normalizing constant $k(\theta)$; see Barndorff-Nielsen (1976), Hyv\"arinen (2007) and Parry {\em et al.} (2012).  The Hyv\"arinen score estimator is an unbiased estimator  for the parameter $\theta$ and its variance coincides with the inverse of the Godambe information $G(\theta)^{-1}\propto Var\left(a'(X)\right)$; see Mameli and Ventura (2015). Thus, $\pi_G(\theta)\propto Var(a'(X))^{-1/2}$, while the Jeffreys prior is $\pi_J(\theta)\propto (\partial ^2 k(\theta)/ \partial \theta^2)^{1/2}$ which requires the knowledge of $k(\theta)$. In models in which $k(\theta)$ is difficult to evaluate,  the Jeffreys prior is not available.


\section{Examples}
In many applications, classical likelihood based methods may be infeasible, for example in models with complex dependency structure, or when robustness with respect to data or to model misspecification is required. In this section,  we consider the analysis of these statistical issues from a scoring rule viewpoint. In particular, we illustrate  the behaviour of the SR-posterior distribution and of the corresponding reference prior by means of three examples. In these examples different non-informative priors are also considered. MCMC methods based on data augmentation, such as the Gibbs sampler, seem difficult to adapt in this setting. The reason is that, in general, the proposed prior and the associated scoring rule may not be conjugate and, the full conditionals may not correspond to known densities. Therefore, except for Example 5.1 in which different posteriors are computed by numerical integration, the posterior distributions are computed via the classical Metropolis-Hastings algorithm using a multivariate normal proposal. The {\tt R} code for fully reproducing the examples of this section can be found in the Electronic Supplementary Material.


\subsection{Hyv\"arinen scoring rule for the von Mises-Fisher distribution}
The first example deals with the study of directional models (Mardia and Jupp, 2000). These models arise naturally in many areas of scientific research, such as meteorology for wind direction, biology for animal movement, in the investigation of biological processes and many others fields. Inference for directional models is difficult because typically the density function contains an intractable normalization constant, which cannot be explicitly computed in closed form. In this setting and to avoid the issue of the intractable normalising constant, Mardia {\em et al.} (2016) propose to use the Hyv\"arinen scoring rule. An important feature of this score is related to the homogeneity property for which the distribution needs to be known up to the normalization constant, thus avoiding the calculation of the density function.

Let us consider the von Mises-Fisher density, which is a directional distribution defined on the unit sphere ${\cal S}_{q-1}\subset\mathbb{R}^q$ given by 
\begin{equation*}
p(z|\kappa)\propto \exp(-\kappa \mu_0^Tz), \quad z\in {\cal S}_{q-1},
\end{equation*}
with $\kappa\in \mathbb{R}^+$ scalar concentration parameter and $\mu_0=(\cos \theta_0, \sin \theta_0)$, with $\theta_0 \in \mathbb{R}$ known.  When $q=2$ and the data are represented in polar coordinates $(z_{h1}, z_{h2}) = (\cos{\theta_h}, \sin{\theta_h})$, $h=1,\ldots,n$, the Hyv\"arinen estimator for $\kappa$ can be derived in closed form and it is given by
\begin{equation}
\tilde{\kappa} = \frac{2\sqrt{\bar{R}^2(1 + \bar{R}_2^2) + 2(\bar{C}^2 - \bar{S}^2)\bar{C}_2 + 4\bar{C}\bar{S}\bar{S}_2}}{(1 - \bar{R}_2^2)},
\end{equation}
where $\bar{C} = \frac{1}{n}\sum_{h=1}^n\cos(\theta_h)\,,\bar{S} = \frac{1}{n}\sum_{h=1}^n\sin(\theta_h)$, $\bar{C}_2 = \frac{1}{n}\sum_{h=1}^n\cos(2\theta_h)$, $\bar{S}_2 = \frac{1}{n}\sum_{h=1}^n\sin(2\theta_h)$ and $\bar{R}_2 = \sqrt{\bar{C}^2_2+ \bar{S}^2_2}.$

\begin{figure}[htbp]
\centering
\includegraphics[scale=0.7]{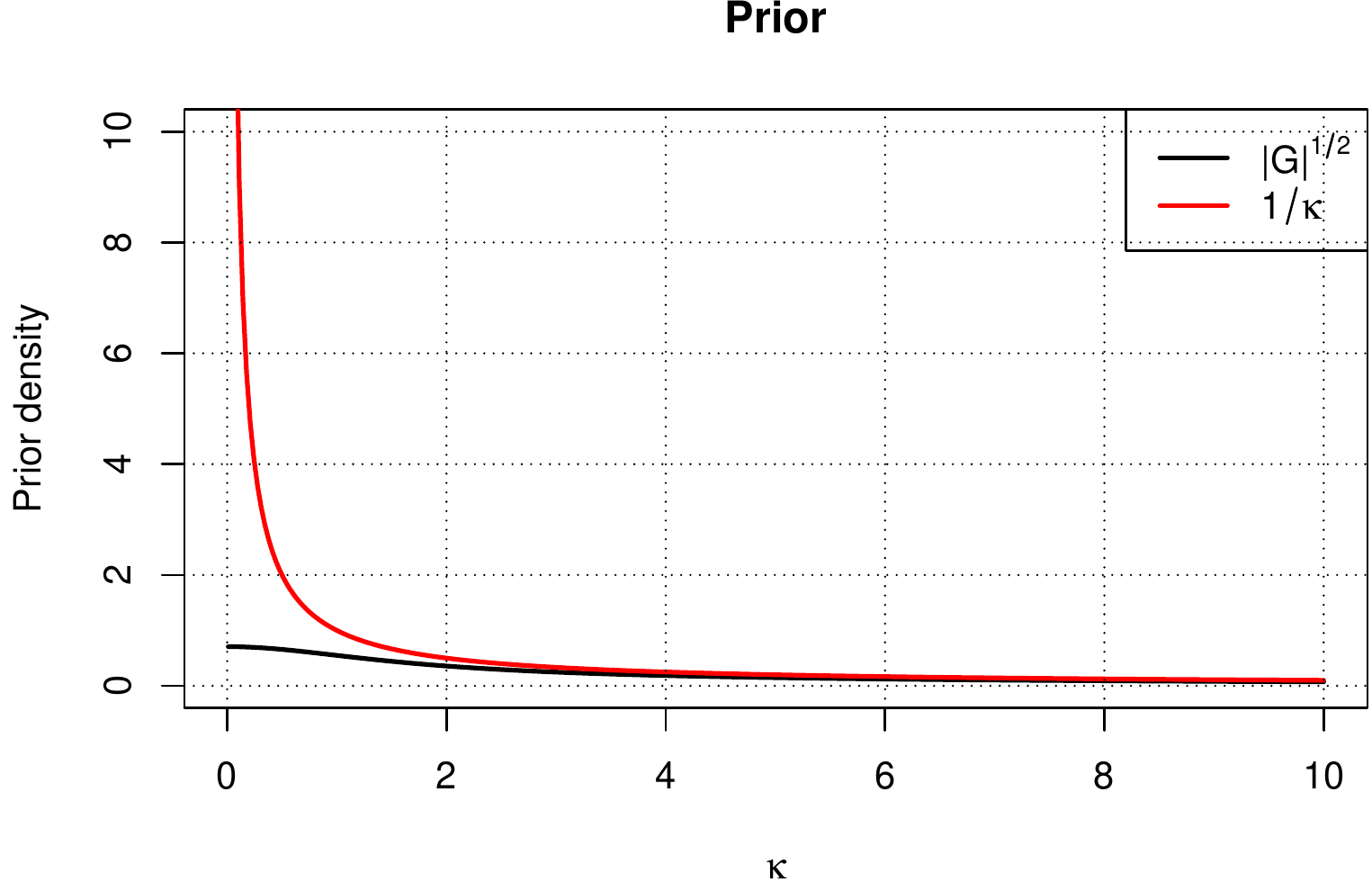}
\caption{Priors $\pi_G(\kappa)\propto |G(\kappa)|^{1/2}$ and \(\pi(\kappa)\propto 1/\kappa\).}
\label{fig2}
\end{figure}

For $\theta_0=0$, Mardia {\em et al.} (2016) show that the asymptotic variance of $\tilde{\kappa} $ is 
$$
\frac{V(\kappa)}{n} = \frac{\kappa\left[2\kappa-3A_1(\kappa)\right]}{nA_1^2(\kappa)},
$$ 
with $A_{1}(\kappa)=I_{1}(\kappa)/I_{0}(\kappa)$, where $I_1$ and $I_0$ are the modified Bessel functions of order 0 and 1, respectively.
The reference prior (\ref{nostraprior}) is thus
\[
\pi_G(\kappa)\propto \sqrt{\frac{A_1^2(\kappa)}{\kappa\left[2\kappa-3A_1(\kappa)\right]}}.
\]
Figure \ref{fig2} shows the plot of the reference prior $\pi_G(\kappa)$ and, for comparison, it also presents the classical non-informative prior \(\pi(\kappa)\propto 1/\kappa\) for the scale parameter. It can be seen from the plot that $\pi_G(\kappa)$ puts finite mass at 0, leading to more appropriate inference when the true parameter values are close to the boundary of the parametric space. Actually, using properties of first order Bessel functions, it can be proved that $\lim_{\kappa\rightarrow 0^+} \pi_G(\kappa)=2^{-1/2}$.

\begin{figure}[htbp]
\centering
\includegraphics[scale=0.8]{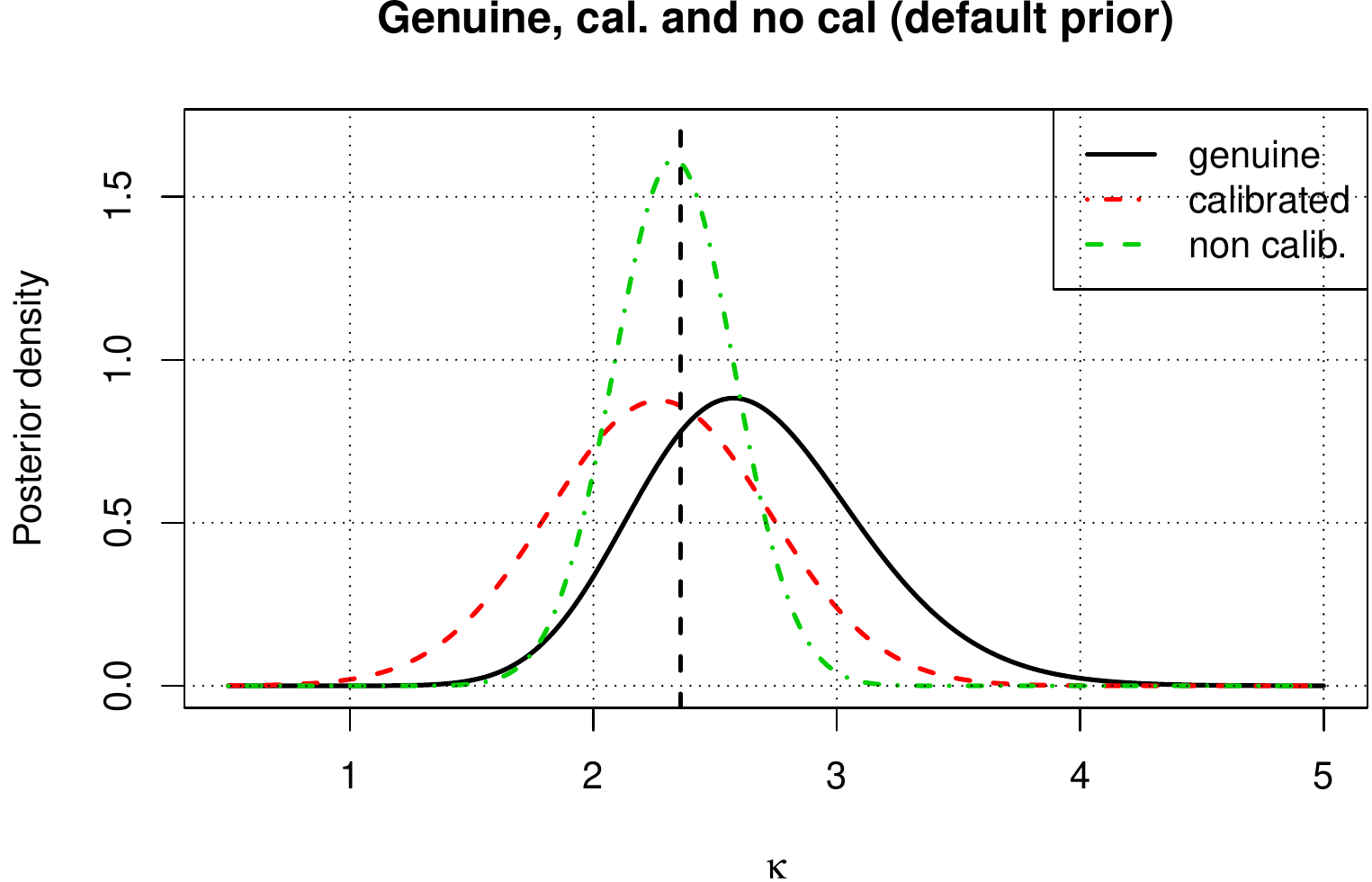}
\caption{Comparison of calibrated SR-posterior vs non calibrated and proper posteriors.}
\label{fig1}
\end{figure}

In order to illustrate that the calibration of the Hyv\"arinen scoring rule is necessary to obtain a posterior distribution (\ref{star}) with the right curvature, consider  a sample of size $n=50$ from the von Mises-Fisher distribution with $(\kappa, \mu_0) = (3,0)$. Hereafter, we take the location parameter $\mu_0$ to be fixed and equal to zero, and consider the problem of estimating $\kappa$. Figure \ref{fig1} compares the full posterior (black line) based on the
likelihood function of the von Mises-Fisher model, with the calibrated (red) and non calibrated (green) SR-posteriors based on the Hyv\"arinen scoring rule. Here, the non-informative prior \(\pi(\kappa)\propto 1/\kappa\) is
used. The vertical line corresponds to the Hyv\"arinen scoring rule estimate of \(\kappa\), i.e.\ $\tilde\kappa=2.36$.
It can be noted that the non calibrated SR-posterior is too narrow compared with the genuine posterior, while the calibrated SR-posterior shows a curvature similar to the genuine posterior.

\begin{figure}[htbp]
\centering
\includegraphics[scale=0.5]{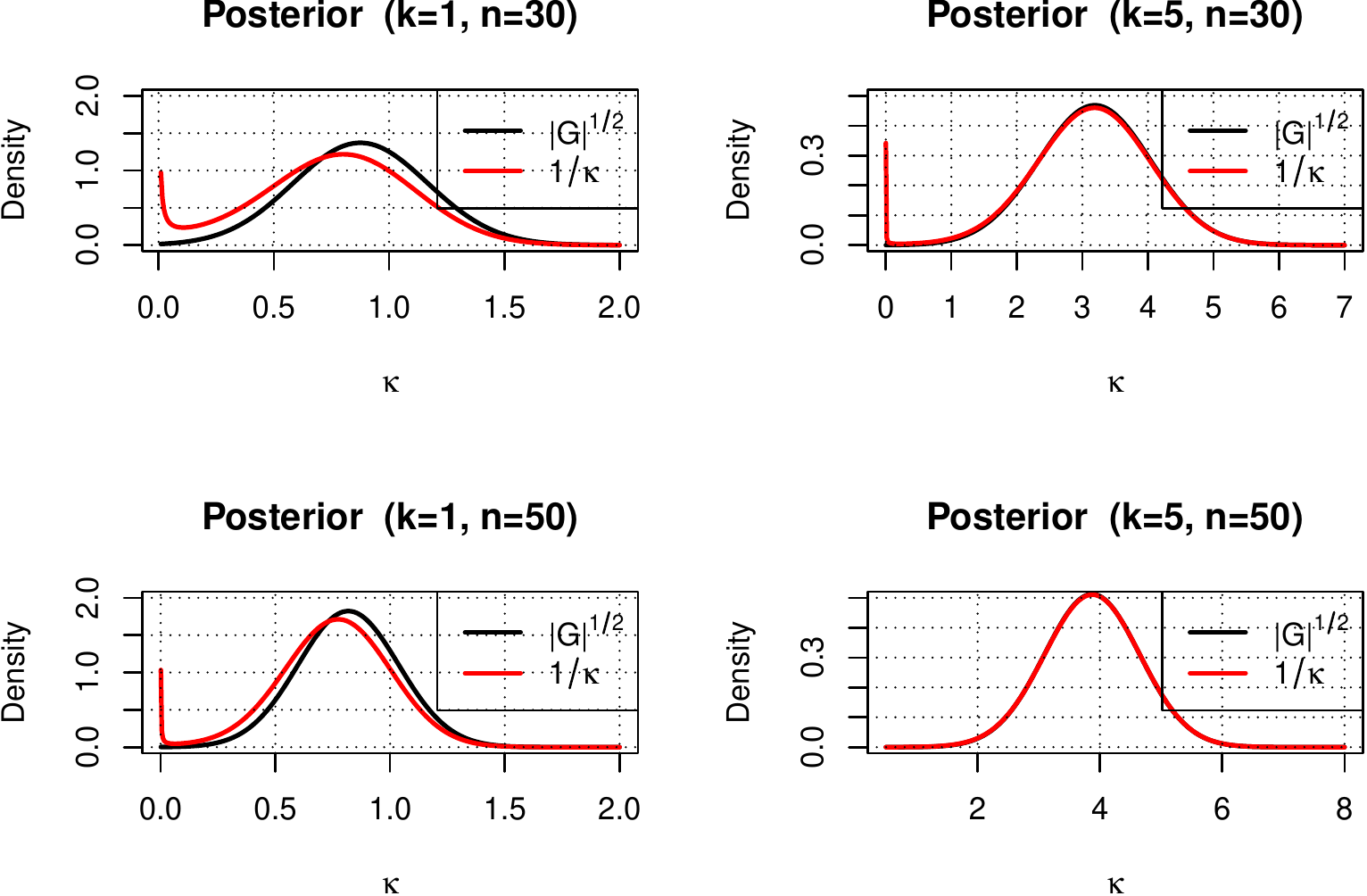}
\includegraphics[scale=0.5]{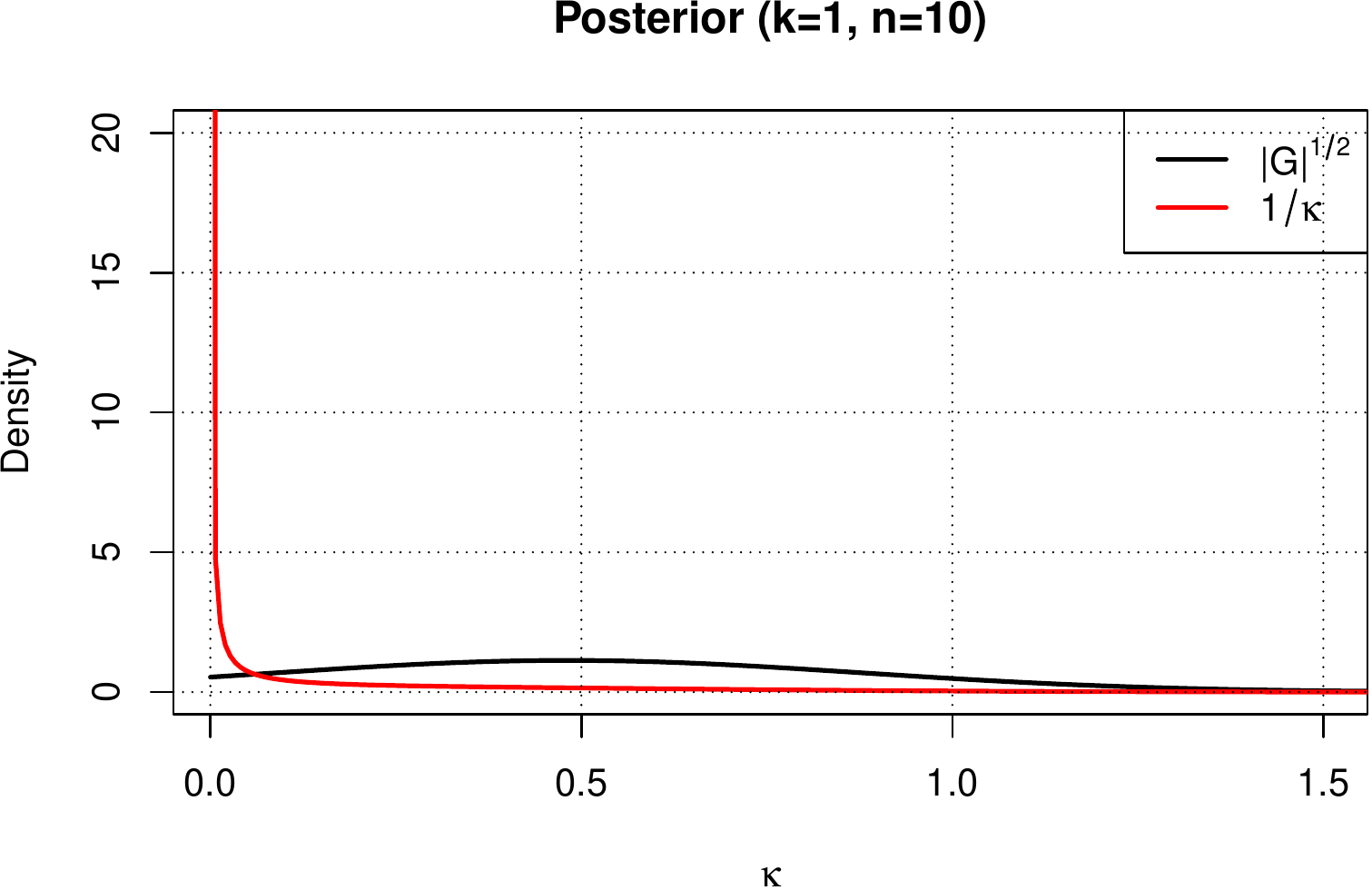}
\caption{SR-posteriors with the $\pi_G(\kappa) \propto |G(\kappa)|^{1/2}$ prior and with the \(\pi(\kappa)\propto 1/\kappa\) prior.}
\label{fig3}
\end{figure}

To compare the behaviour of the SR-posterior distributions, based on the two priors $\pi_G(k)\propto |G|^{1/2}$ and \(\pi(\kappa)\propto 1/\kappa\), two scenarios have been considered and the SR-posteriors with the two priors have been computed. The SR-posteriors are given in Figure \ref{fig3} for different values of the parameters $n$ and $\kappa$. In the first scenario, the true value of \(\kappa\) approaches the boundary and/or the sample size is small. In particular, we consider samples of size $n=10,30,50$ with \(\kappa=1\). It can be noted that for small samples sizes or for the true value of \(\kappa\) near zero, the SR-posterior with \(\pi(\kappa)\propto 1/\kappa\) may not be proper or puts too much mass at zero, i.e.\ it has a vertical asymptote at zero. On the other hand, the SR-posterior with the reference prior $\pi_G(\kappa)$ is proper and unimodal.

In the second scenario, the true value of \(\kappa\) is away from the boundary and the sample size is moderate or large. We consider samples of size \(n=30,50\), with \(\kappa=5\). The SR-posteriors in these cases look very similar, as it can be noted from Figure \ref{fig3}. Indeed, for high values of \(\kappa\) the two SR-posteriors are indistinguishable.


\subsection{Pairwise likelihood for the multivariate equi-correlated normal model}

Let $X$ be a $q-$dimensional random vector with mean $\mu$ and covariance matrix $\Sigma$, with $\Sigma_{rr} = \sigma^2$ and $\Sigma_{rs} = \rho\sigma^2$ for $r\neq s$, with $r,s = 1,\ldots, q$ and $\rho\in (-1/(q-1),1)$. The pairwise log-likelihood for $\theta = (\mu, \sigma^2,\rho)$  is (see Pace {\em et al.}, 2011)
\begin{eqnarray}
S_p (\theta) &=& -\frac{nq(q - 1)}{2}\log{\sigma^2} -\frac{nq(q - 1)}{4}\log{(1 - \rho^2)} - \frac{q - 1 + \rho}{2\sigma^2(1 - \rho^2)}SSW+\nonumber\\&-& \frac{q(q - 1)SSB + nq(q - 1)( \bar{y} - \mu)^2}{2\sigma^2(1 + \rho)},\nonumber
\end{eqnarray}
where $SSW =\sum_{i=1}^n\sum_{r=1}^q(x_{ir} - \bar{x}_i )^2$, $SSB =\sum_{i=1}^n( \bar{x}_i- \bar{x})^2$, $\bar{x}_i =\sum_{r=1}^q x_{ir} /q$ and $\bar{x} =\sum_{i=1}^n\sum_{r=1}^q x_{ir}/(nq)$.

The reference prior (\ref{nostraprior}) is proportional to the square root of the determinant of the Godambe information matrix whose components are given in Pace {\em et al.}\ (2011).

\begin{figure}[htbp]
\centering
\includegraphics[scale=0.8]{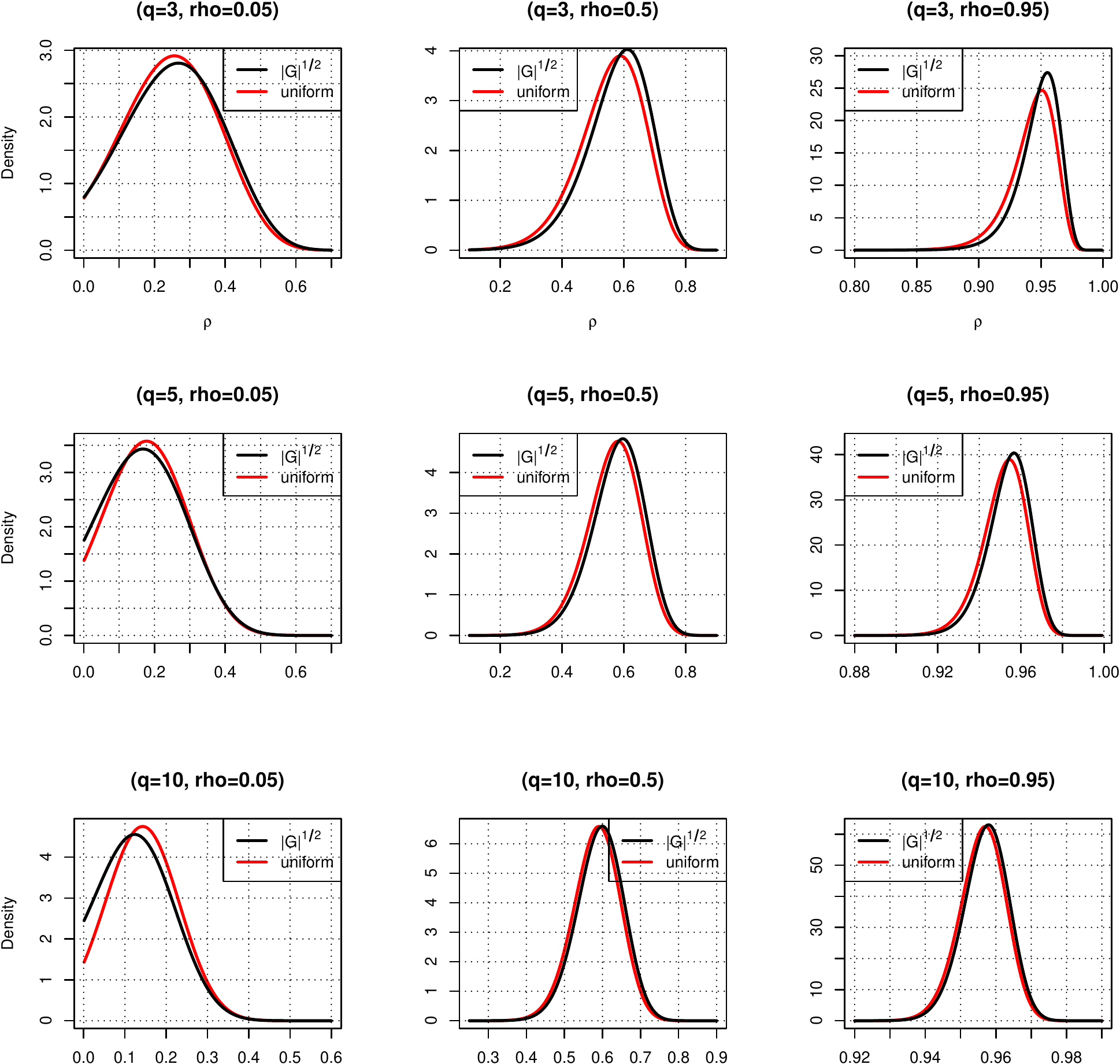}
\caption{SR-posteriors with the $\pi_G(\rho)$ and with the uniform prior in $(0,1)$ for different values of $\rho$ and $q$ where $n=10$, $\sigma^2=1$ and $\mu=0$.}
\label{fig4}
\end{figure}

Consider, first, the simplest situation with $\mu$ and $\sigma^2$ known. The SR-posterior distribution based on the pairwise likelihood with the uniform prior and the proposed reference prior is computed for different values of \(\rho\) and \(q\) (see Figure \ref{fig4}). We notice that the two priors give very similar results when the value of the parameter $\rho$ is away from zero.

Consider now the multi-parametric case. We compare the SR-posteriors based on the pairwise likelihood with the reference prior and with a non-informative prior derived by using two
alternative parametrizations:
\begin{itemize}
\item[(a)] $\theta = (\mu, \sigma, \rho)$, for which we assume the non-informative prior $\pi(\theta) \propto 1/\sigma$.
\item[(b)] $\xi = (\mu, \tau, \kappa)$, with $\tau = \log(\sigma)$ and $\kappa = \text{logit}(\rho)$, for which we assume the flat prior $\pi(\xi)\propto 1$.
\end{itemize}
The need for calibration of the composite likelihood has been studied by Ribatet {\em et al.} (2012); see also Pauli {\em et al.} (2011).
The calibrated SR-posterior distribution has been computed with the parametrization \(\xi\) by using formula (\ref{posterior}), in order to avoid the constraints on $\theta \in \Theta$.

\begin{figure}[htbp]
\centering
\includegraphics[scale=0.7]{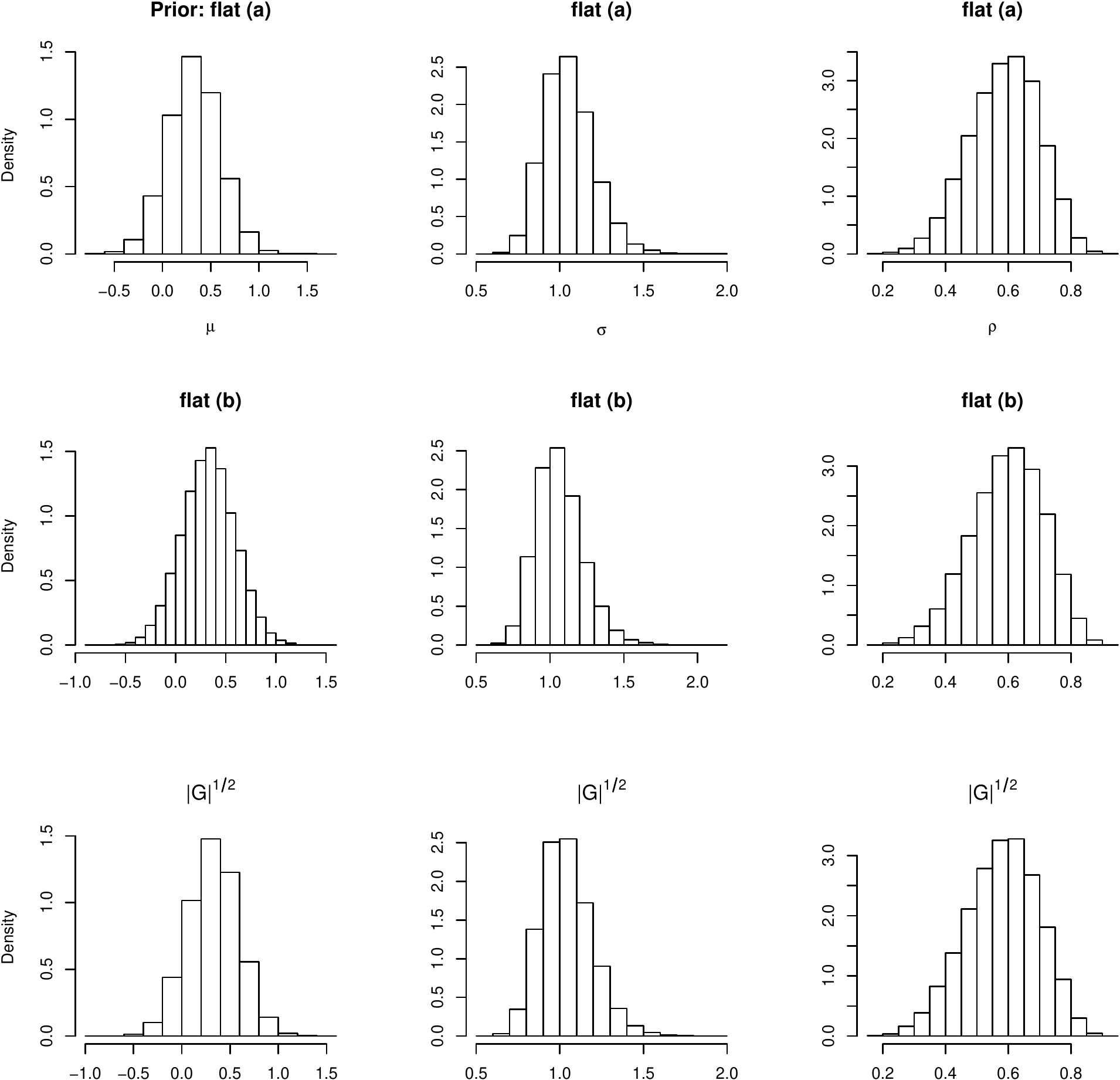}
\caption{SR-posterior with the reference prior $\pi_G(\theta)$ compared with the two alternative flat priors (a) and (b). The data are generated from the equi-correlated model with $n=10$, $q=10$, $\mu=0$, $\sigma^2=1$ and $\rho=0.5$}
\label{fig5}
\end{figure}

As a first example, consider \(n=10\), \(q=10\) and a sample generated under the equi-correlated normal model with \(\theta=(0, 1, 0.5)\). The SR-posterior distributions with the reference prior and the two non-informative priors in cases $(a)$ and $(b)$ are shown in Figure \ref{fig5}. The SR-posterior distributions, described by means of histograms, show no appreciable differences.

\begin{figure}[htbp]
\centering
\includegraphics[scale=0.6]{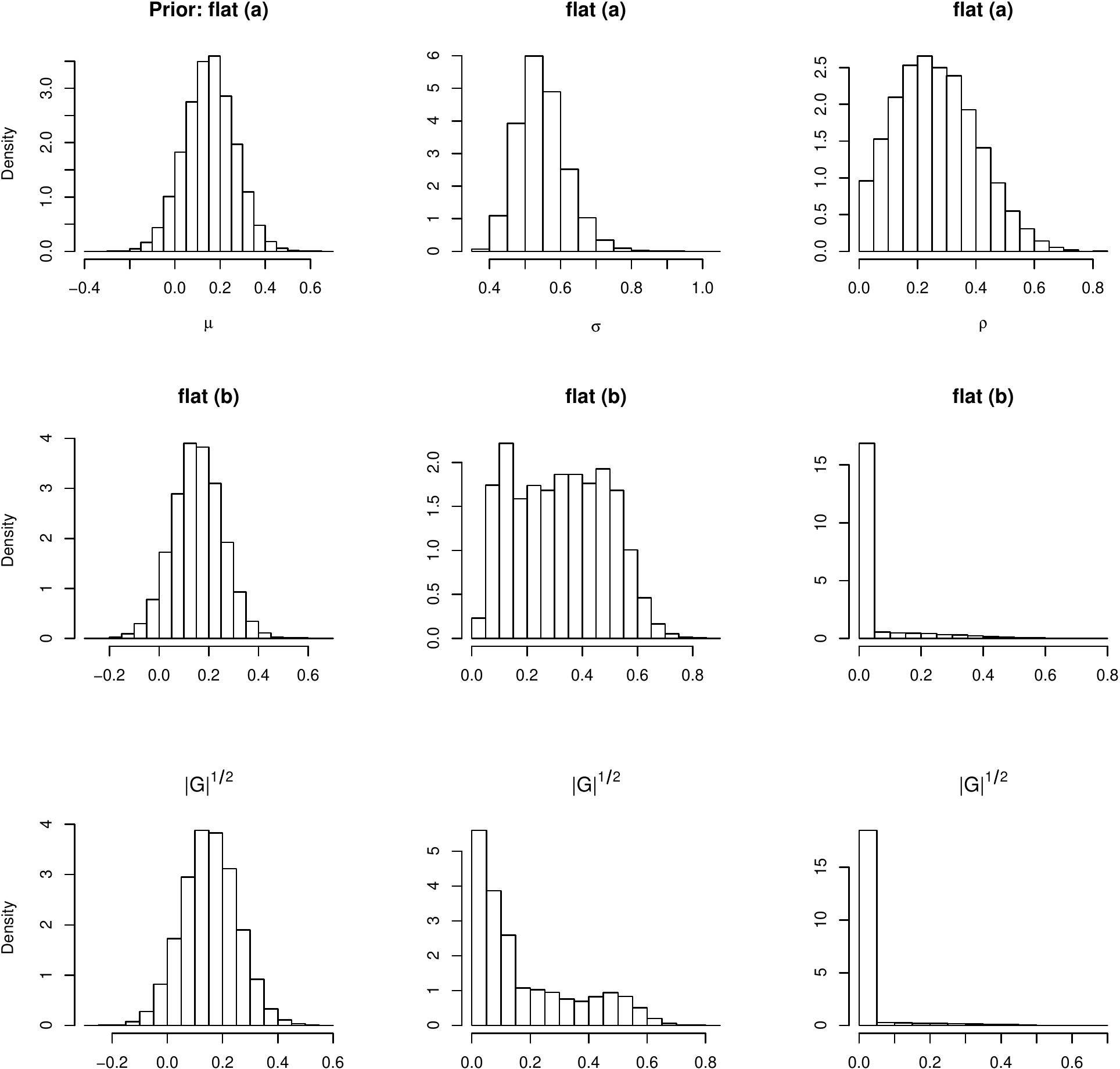}
\caption{SR-posterior with the reference prior $\pi_G(\theta)$ (third row) compared with the two alternative flat priors (a) (first row) and (b) (second row). The data are generated from the equi-correlated model with $n=10$, $q=4$ $\mu=0$, $\sigma^2=0.5$ and $\rho=0.1$}
\label{fig6}
\end{figure}

As a second example, consider the more problematic scenario with \(n=10\), \(q=4\) and data generated under the equi-correlated normal model with \(\theta=(0, 0.5, 0.1)\). The SR-posteriors with the reference prior and the two non-informative priors are shown in Figure \ref{fig6}.

It can be noted that a flat prior on \(\theta\) may not necessarily be flat also on \(\xi\), at least not for \(\sigma\) and \(\rho\). Indeed, the uniform prior on \(\rho\) implies a standard logistic prior  on \(\kappa\), which is quite informative. This explains the ``regularised" behaviour of the marginal posteriors of $\rho$ and $\sigma$ in case $(a)$. On the other hand, the proposed reference prior behaves quite similarly to the uniform improper prior \(\pi(\xi)\propto 1\).


\subsection{Tsallis scoring rule for linear regression models}
The third example is related to robustness procedures, which are more stable and reliable than their classical likelihood-based counterparts under misspecification. As it is well known, the ordinary Bayes estimator based on the classical based likelihood posterior density lacks robustness. Gosh and Basu (2016) studied the robustness properties of the posterior distribution based on the Tsallis scoring rule. There is a trade-off between robustness ad efficiency in the Tsallis scoring rule: small values of $\gamma$ in \eqref{eq:tsallisscore} entails to a small loss in efficiency but a reduced robustness; while large values of $\gamma$ are associated to great robustness but low efficiency. Here, we consider a linear regression model to demonstrate the robustness performance of the Tsallis scoring rule and of the derived prior.\\
Consider the linear regression model
\begin{equation*}
y =  X \beta+ \sigma \epsilon,
\end{equation*}
where $X$ is a fixed $n \times p$ matrix, $\beta \in \mathbb{R}^p$ ($p \geq 1$) is the vector of unknown regression coefficients, $\sigma > 0$ is a scale parameter, and $\epsilon$ is an $n-$dimensional vector of random errors from a standard normal distribution. 

Let $\theta=(\beta,\sigma^2)$. The Tsallis total empirical score in this case assumes the following form (Ghosh and Basu, 2013)
\begin{equation*}
S_T(\theta)= \frac{\gamma}{(2\pi\sigma^2)^{\frac{\gamma-1}{2}}}\sum_{i=1}^n e^{-\frac{(\gamma-1)}{2\sigma^2}(y_i-x_i^T\beta)^2}-\frac{n (\gamma-1)}{\sqrt{\gamma}(2\pi\sigma^2)^{(\gamma-1)/2}}.
\end{equation*}
The asymptotic variance of the Tsallis estimator of $\theta$ is given in Ghosh and Basu (2013). In particular, the asymptotic distribution of $(X^TX)^{1/2}(\tilde{\beta}-\beta)$ is a $N_p(0,v^{\beta}_{\gamma}I_p)$, with $v^{\beta}_{\gamma}=\sigma^2\left( 1+\frac{(\gamma-1)^2}{2\gamma-1}\right)^{3/2}$, while $\sqrt{n}(\tilde{\sigma}^2 - \sigma^2)$ follows a normal distribution with mean 0 and variance $v^e_{\gamma}$, where 
$$
v^{e}_{\gamma}= \frac{4\sigma^4}{(2+(\gamma-1)^2)^2}\left(2(1+2(\gamma-1)^2)\left(1+\frac{(\gamma-1)^2}{2\gamma-1}\right)^{5/2}-(\gamma-1)^2\gamma^2\right).
$$ 
Since the asymptotic distributions of $\tilde{\beta}$ and $\tilde{\sigma}^2$ are independent, the proposed reference prior is proportional to the square root of the determinant of the Godambe information, i.e.\
\begin{eqnarray}
\pi_G(\theta)\propto \left(v^{\beta}_{\gamma}v^{e}_{\gamma}\right)^{-1/2}.
\label{prioreg}
\end{eqnarray}
Therefore, a prior for $\beta$ is $\pi_G(\beta)\propto \left(v^{\beta}_{\gamma}\right)^{-1/2}$, while a prior for $\sigma^2$ is 
$\pi_G(\sigma^2)\propto \left(v^{e}_{\gamma}\right)^{-1/2}.$

\begin{figure}
\centering
\includegraphics[scale=0.8]{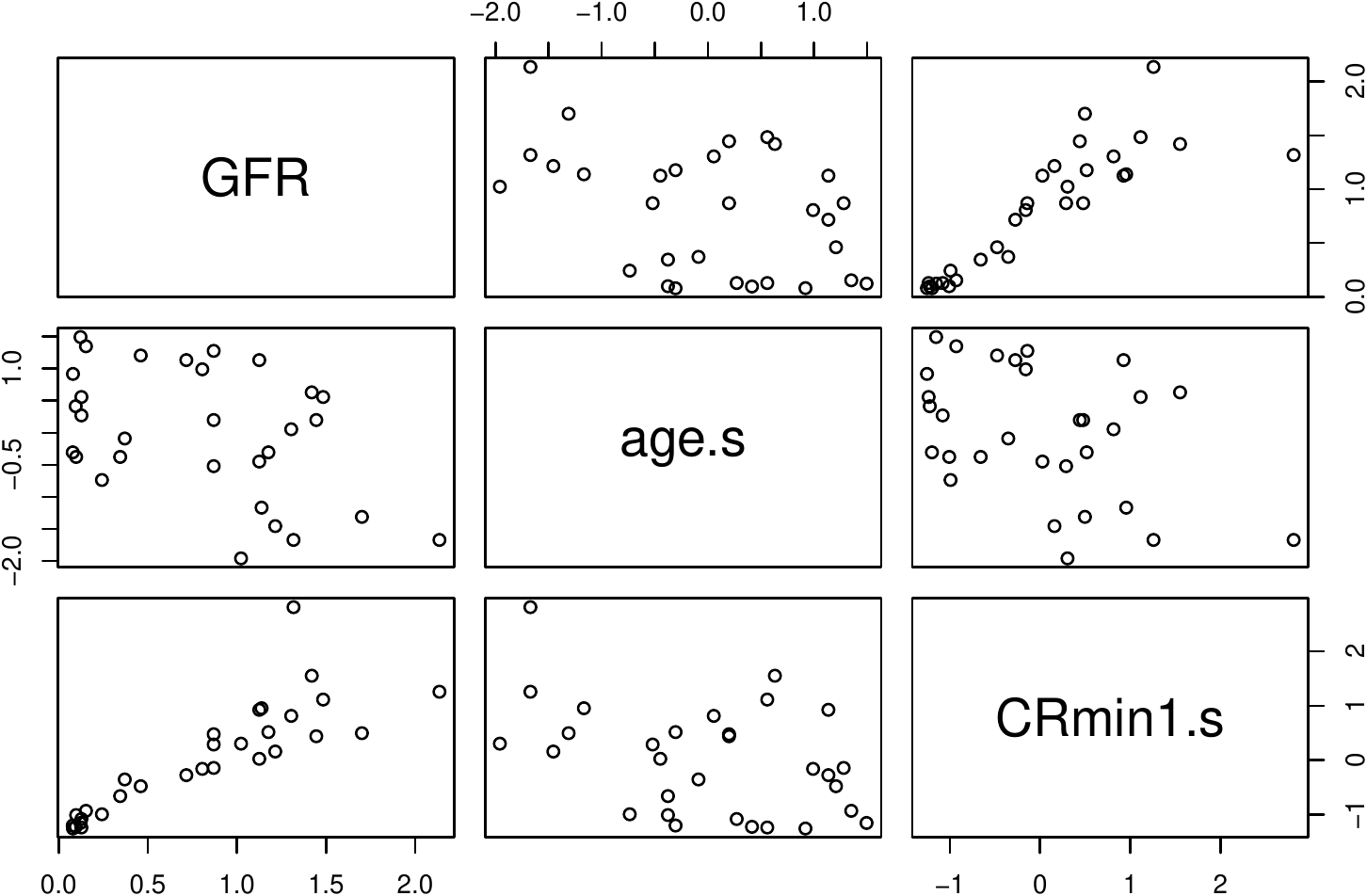}
\caption{Scatterplot diagram of GFR data.}
\label{scatter}
\end{figure}

In order to illustrate the behaviour of the Tsallis scoring rule in the context of the linear regression model, we consider the GFR dataset (Heritier {\em et al.}, 2009), which contains measurements of the glomerular filtration rate (GFR) and serum creatinine (CR) on $n=30$ subjects. The GFR is the volume of fluid filtered from the renal glomerular capillaries into the Bowmans capsule per unit of time (typically in millilitres per minute) and, clinically, it is often used to determine renal function. Its estimation, when not measured, is of clinical importance and several techniques are used for that purpose. One of them is based on CR, an endogenous molecule, synthesized in the body, which is freely filtered by the glomerulus (but also secreted by the renal tubules in very small amounts).  Several models have been proposed in the literature to explain the logarithm of GFR as a function of CR. Here, following Heritier {\em et al.} (2009), we consider a model for GFR based on CR$^{-1}$ and AGE, i.e.
$$
\text{GFR} = \beta_0 + \beta_1 \,  \frac{1}{\text{CR}} + \beta_2 \, \text{AGE} + \varepsilon.
$$
Both covariates are scaled to have mean zero and unit variance, whereas the response is scaled to have unit variance. The data are illustrated in  Figure \ref{scatter}: note that there are some observations which look like outliers. The need and consequence to resort to robust procedures to deal with these data have been highlighted in several papers; see, among others, Heritier {\em et al.} (2009) and Farcomeni and Ventura (2012).

\begin{figure}
\centering
\includegraphics[scale=0.7]{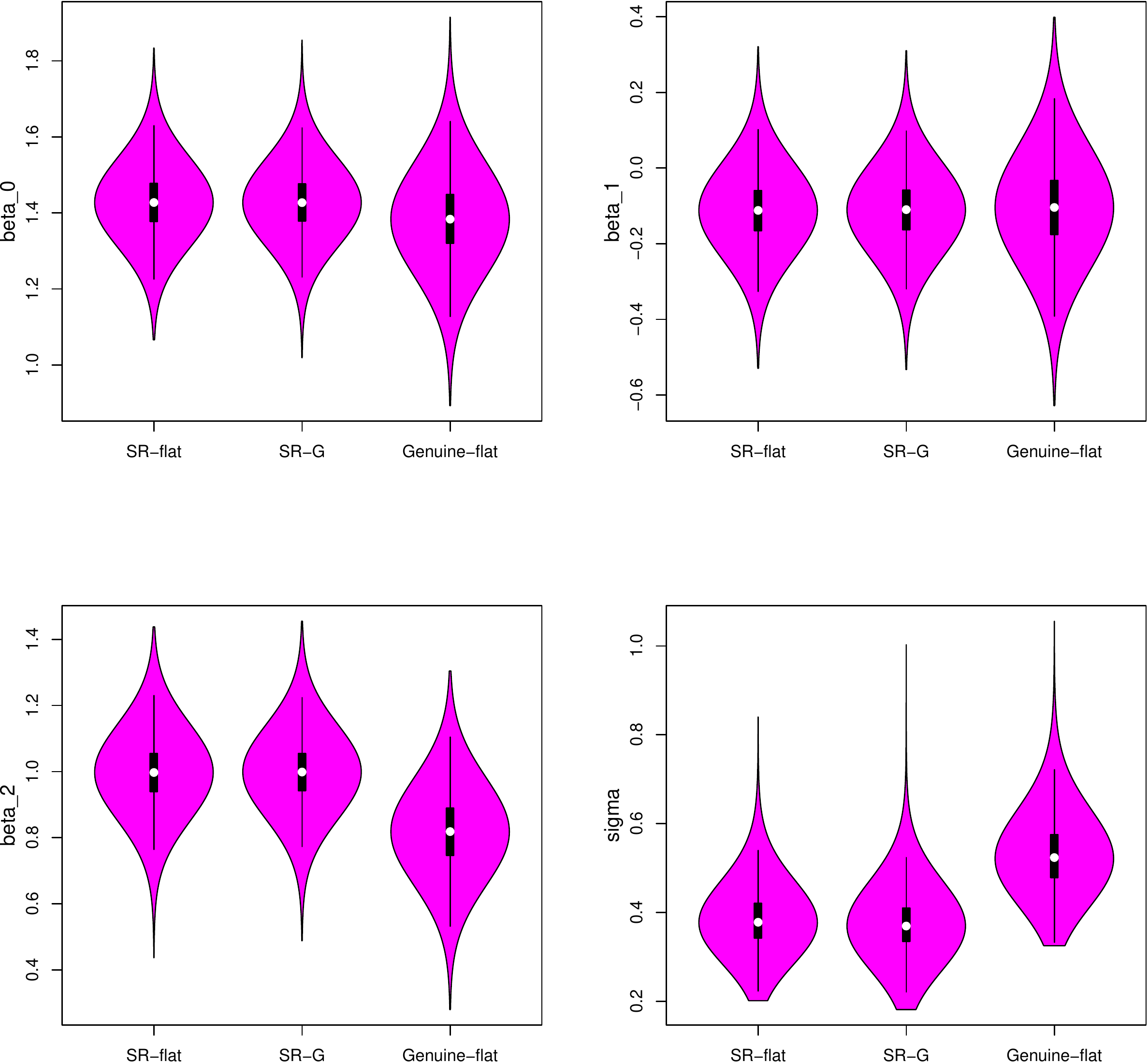}
\caption{SR-posteriors based on the Tsallis score with $\gamma=1.25$ (with the flat or the reference prior) vs the genuine posterior for the GFR dataset.}
\label{reg1}
\end{figure}

For the parameter \(\theta = (\beta, \log\sigma)\) we assume both the usual non-informative flat prior \(\pi(\theta)\propto 1\) and the proposed reference prior (\ref{prioreg}) in the SR-posterior distribution, while the classical posterior distribution is based on the non-informative flat prior. Figure \ref{reg1} gives the violin plots of the marginal SR-posterior distributions based on the Tsallis score, with \(\gamma = 1.25\),  and of the classical posterior distribution. From Figure \ref{reg1} it can be noted that  the proposed reference prior behaves similarly to the non-informative prior. Moreover, the classical marginal posterior distribution shows in general heavier tails and, in particular, the robust and the classical posterior distributions give different inferences on $\beta_2$ and $\sigma$.

\begin{figure}
\centering
\includegraphics[scale=0.5]{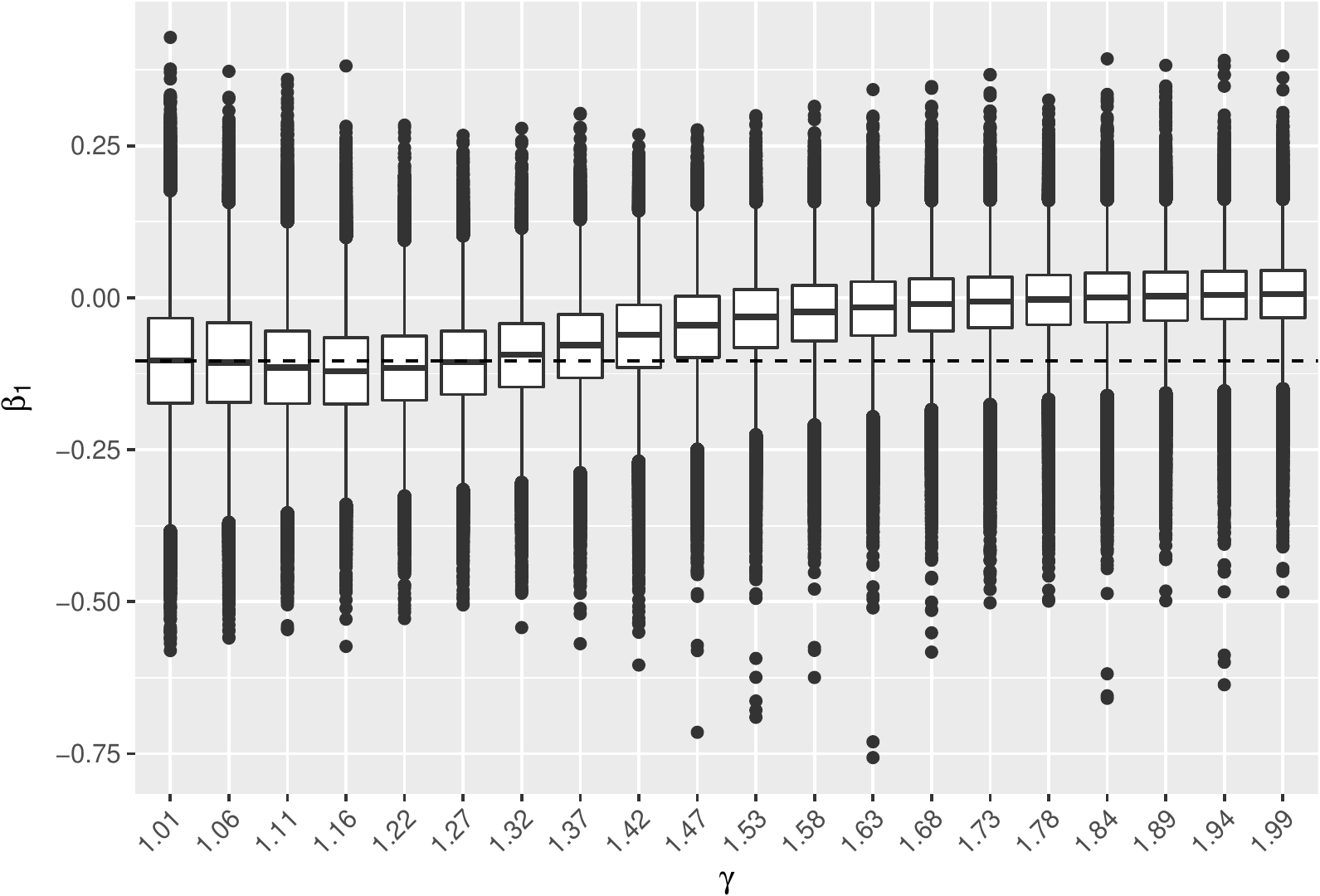}
\includegraphics[scale=0.5]{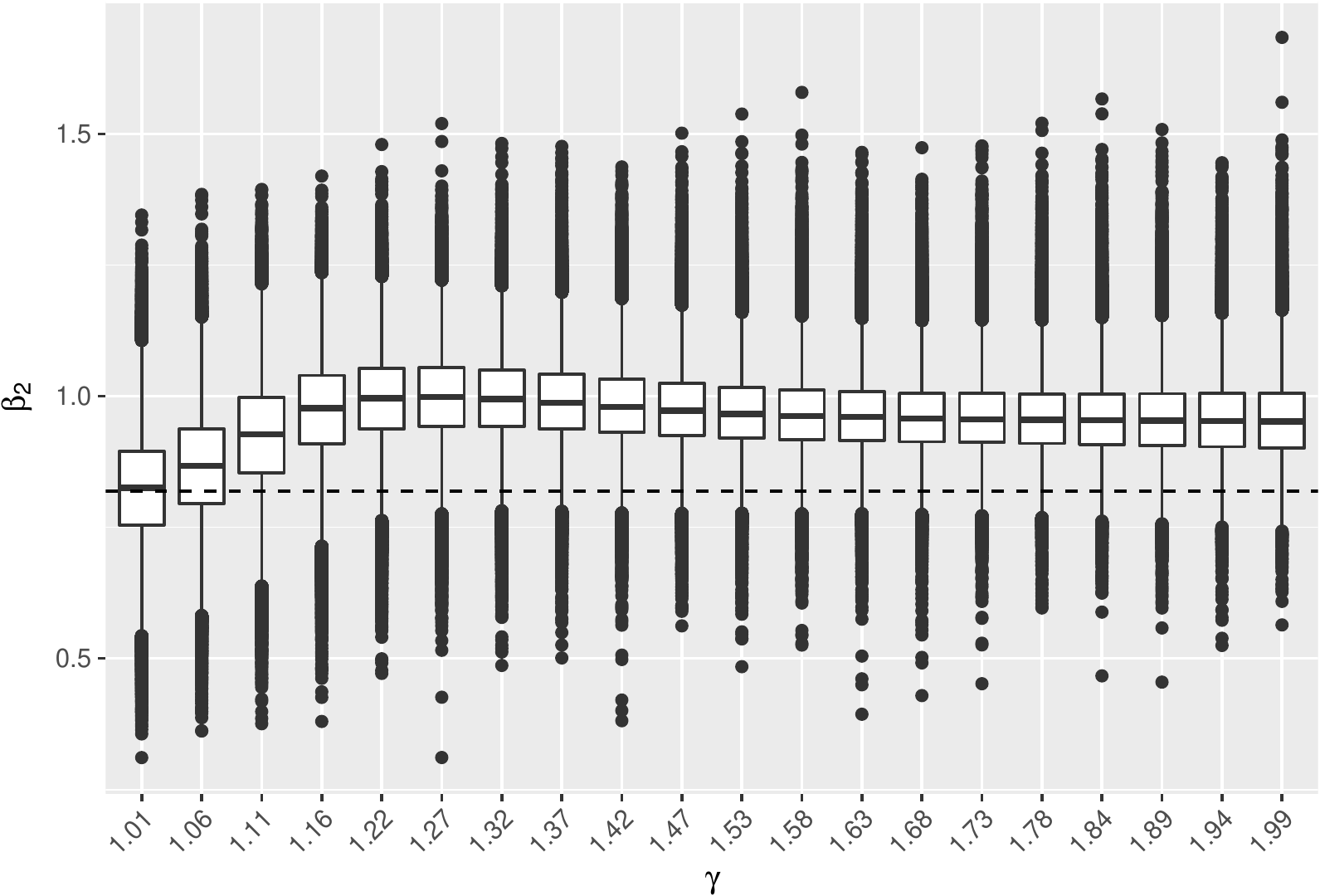}
\includegraphics[scale=0.5]{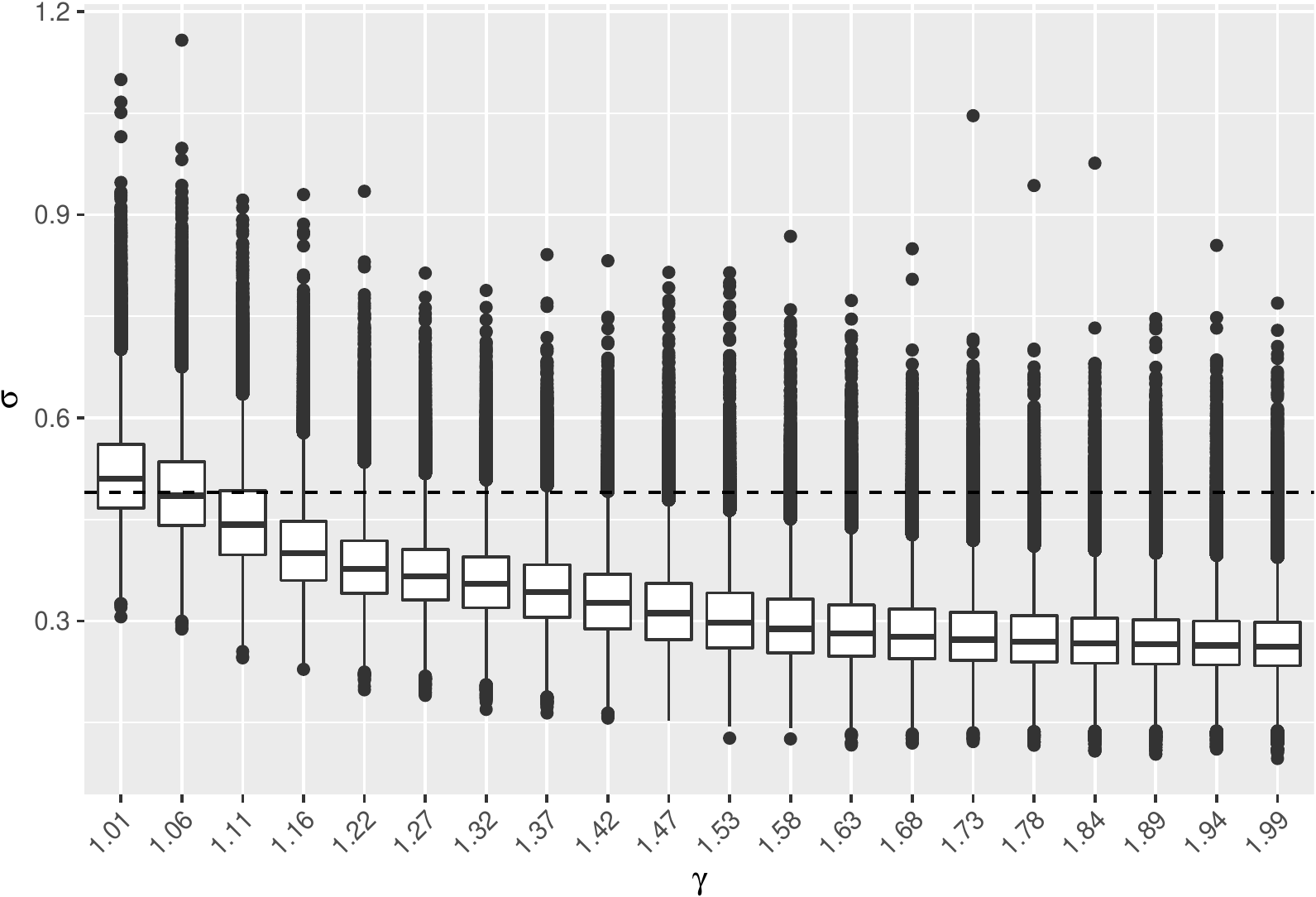}
\includegraphics[scale=0.5]{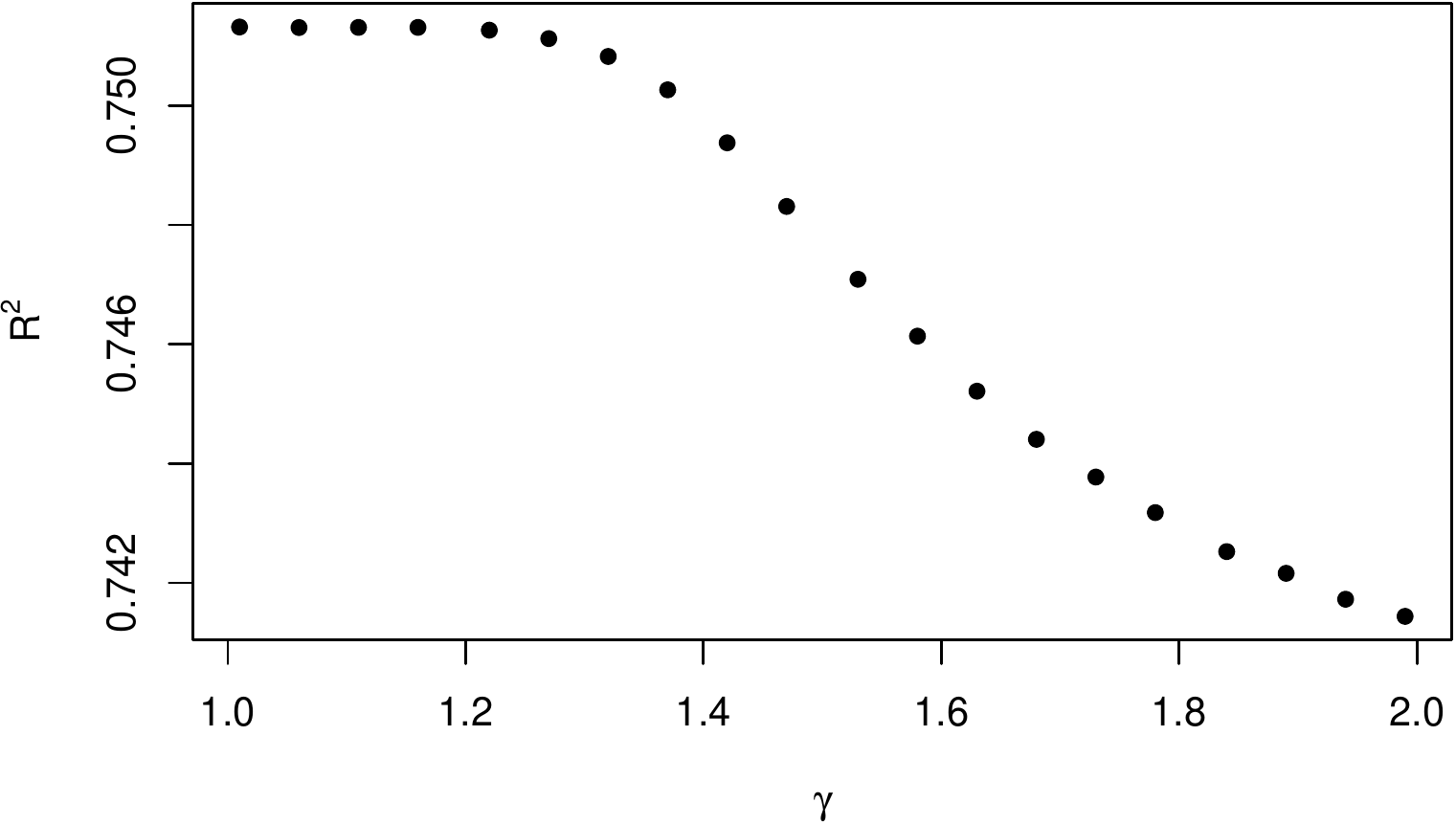}
\caption{Marginal SR-posteriors based on the Tsallis score, for $ \beta_1, \beta_2$ and $\sigma$ with the reference prior, and values of $R^2$ for varying values of $\gamma$.}
\label{figa}
\end{figure}

\newpage
Figure \ref{figa} gives useful monitoring plots in robustness studies. In particular, these plots illustrate the SR-posterior distributions based on the Tsallis score with reference prior as a function of $\gamma$. This approach (see, e.g., Riani {\em et al.}, 2014)  provides tools for gaining knowledge and better understanding of the properties of robust procedures. The horizontal lines in Figure \ref{figa}  correspond to the posterior mode of the genuine posterior based on the flat prior. 
The value $\gamma=1$ corresponds to complete overlap of the two methods. As far as $\gamma$ increases, robustness is achieved although at the price of losing efficiency at the central model; the value $\gamma=1.25$ gives approximately 0.95\% efficiency under the normal distribution. 

As a final remark, note that the Tsallis scoring rule provides a cogent framework for dealing with robust procedures in the Bayesian framework.


\section{Conclusions}

In this paper, we discuss the use of scoring rules in order to compute a posterior distribution, useful to deal with complex models or if robustness with respect to data or to model misspecification is required. Indeed, scoring rules provide a flexible and robust way of combining data-driven information with prior distributions, either subjective or non-informative. 

One should devise a scoring rule that properly captures the structure of the data, otherwise the resulting posterior inferences are not reliable. In this respect, the result of Section 3 shows that a suitably calibrated SR-posterior distribution is, up to first order,  
normally distributed with the same asymptotic variance of the scoring rule estimator.

When dealing with default priors, in Section 4 reference priors for a vector parameter based on maximising $\alpha-$divergences are discussed in the framework of scoring rules. We show that, for $0 \leq  |\alpha| < 1$, the result is a Jeffreys-type prior that is proportional to the square root of the determinant of the Godambe information matrix.

A possible direction of further research is to extend the proposed methodology considering the class of monotone and regular divergences which is a broad family of divergences asymptotically equivalent to $\alpha-$divergences; see Corcuera and Giummol\`e (1998).

\section*{Appendix}

\subsection*{Proof of Theorem \ref{uno}}

\begin{proof}
The posterior (\ref{star}) can equivalently be written as
\[
\pi_{SR} (\theta|x)=\frac{\pi(\theta) \exp{\{-S (\theta^*)+S(\tilde\theta)\}}}
{\int_\Theta \pi(\theta) \exp{\{-S (\theta^*)+S(\tilde\theta)\}} d\theta},
\]
with $\theta^*=\tilde\theta+C(\theta-\tilde\theta)$ and $C$ fixed such that $C^TK(\theta)C=G(\theta)$.

Let $w=(w^1,\dots,w^d)^T=n^{1/2}(\theta-\tilde\theta)$. Then, $\theta=\tilde\theta+n^{-1/2}w$ and $\theta^*=\tilde\theta+n^{-1/2}Cw$. A posterior for $w$ is
\[
\pi_{SR} (w|x)=\frac{b(w,x)}{\int b(w,x) dw},
\]
with
\[
b(w,x)= \pi(\tilde\theta+n^{-1/2}w)   \exp{\{-S (\tilde\theta+n^{-1/2}Cw)+S(\tilde\theta)\}}.
\]
Now
\[
\pi(\tilde\theta+n^{-1/2}w)=
\tilde\pi\left(1+n^{-1/2}R_1(w)+\frac{1}{2}n^{-1}R_2(w)\right)+O_p(n^{-3/2}),
\]
with
$$
R_1(w)=\frac{\tilde\pi_i}{\tilde\pi}w^i
\quad\quad \mbox{ and }\quad\quad   
R_2(w)=\frac{\tilde\pi_{ij}}{\tilde\pi}w^{ij},
$$
where $w^{ij\dots}=w^iw^j\dots$ is a product of components of $w$.
Moreover,
\begin{eqnarray}
-S (\tilde\theta+n^{-1/2}Cw)+S(\tilde\theta)&=&
-n^{-1}\left(\frac{1}{2}(Cw)^{ij}\tilde S_{ij}
+\frac{1}{6}n^{-1/2}(Cw)^{ijk}\tilde S_{ijk}\right. \nonumber\\
&&\left. +\frac{1}{24}n^{-1}(Cw)^{ijkh}\tilde S_{ijkh}\right)+O_p(n^{-3/2}) \nonumber\\
&=&-\frac{1}{2}n^{-1}w^T\left(C^T\frac{\partial^2\tilde S}{\partial\theta\partial\theta^T}C\right)w-\frac{1}{6}n^{-1/2}R_3(w)\nonumber\\
&&-\frac{1}{24}n^{-1}R_4(w) +O_p(n^{-3/2})\nonumber \\
&=&-\frac{1}{2}w^T\tilde H w-\frac{1}{6}n^{-1/2}R_3(w)-\frac{1}{24}n^{-1}R_4(w) +O_p(n^{-3/2}),\quad\quad
\end{eqnarray}
where 
$$R_3(w)=n^{-1}(Cw)^{ijk}\tilde S_{ijk} \quad\quad\mbox{   and   }\quad\quad
R_4(w)=n^{-1}(Cw)^{ijkh}\tilde S_{ijkh}.$$

The numerator in Bayes formula can thus be written as
\begin{eqnarray}
\label{num}
\lefteqn{b(w,x)=\tilde\pi\left(
1+n^{-1/2}R_1(w)
+\frac{1}{2}n^{-1} R_2(w)\right)}\nonumber\\
&&\exp{\left\{ -\frac{1}{2}w^T\tilde H w \right\}}\left(1-\frac{1}{6}n^{-1/2}R_3(w)-\frac{1}{24}n^{-1}R_4(w)
+\frac{1}{72}n^{-1}R_3(w)^2\right) +O_p(n^{-3/2})\nonumber\\
&=& 
\tilde\pi \exp{\left\{ -\frac{1}{2}w^T\tilde H w \right\}}\left[
1+n^{-1/2}\left( R_1(w)-\frac{1}{6}R_3(w)\right)\right.\nonumber\\
&&\left.+n^{-1}\left( \frac{1}{2}R_2(w) -\frac{1}{6}R_1(w)R_3(w)-\frac{1}{24}R_4(w)+ \frac{1}{72}R_3(w)^2
\right)
\right] +O_p(n^{-3/2}).
\end{eqnarray}

The denominator can be approximated using the moments of the $d-$variate normal distribution $N_d(0,\tilde H^{-1})$.  We have that
\begin{eqnarray}
\label{denom}
\int b(w,x) dw&=&\tilde\pi (2\pi)^{d/2}|\tilde H|^{-1/2}
\left[1+n^{-1}\left( \frac{1}{2}E(R_2(W)) -\frac{1}{6}E(R_1(W)R_3(W))\right.\right.\nonumber\\ 
&&\left.\left.-\frac{1}{24}E(R_4(W))+ \frac{1}{72}E(R_3(W)^2)
\right)
\right] + o(n^{-1}),
\end{eqnarray}
being $E(R_1(W))$ and $E(R_3(W))$ both related to odd moments and thus equal to zero.
Now,
$$
E(R_2(W))= \frac{\tilde\pi_{ij}}{\tilde\pi} \tilde h^{ij}  ,
\quad\quad
E(R_1(W)R_3(W))=  3\frac{\tilde\pi_{i}}{\tilde\pi} \frac{\tilde S_{jkh}}{n}c_{jr}c_{ks}c_{ht}\tilde h^{ir}\tilde h^{st}  ,
$$
$$
E(R_4(W))= 3 \frac{\tilde S_{ijkh}}{n}c_{ir}c_{js}c_{kt}c_{hu}\tilde h^{rs}\tilde h^{tu}   
$$
and
\begin{eqnarray*}
E(R_3(W)^2)&=&  \frac{\tilde S_{ijk}\tilde S_{rst}}{n^2}(9c_{ia}c_{jb}c_{kc}c_{rd}c_{se}c_{tf}\tilde h^{ab}\tilde h^{cd}\tilde h^{ef}\\
&&+6c_{ia}c_{rb}c_{jc}c_{sd}c_{ke}c_{tf}\tilde h^{ab}\tilde h^{cd}\tilde h^{ef}) .
\end{eqnarray*}

Putting together expressions (\ref{num}) and (\ref{denom}), we finally obtain the result with
$$ A_1(w)=R_1(w)-\frac{1}{6}R_3(w)$$
and
\begin{eqnarray*} 
A_2(w)&=&\frac{1}{2}\left(R_2(w)-E(R_2(W))\right)
-\frac{1}{6}\left(R_1(w)R_3(w)-E(R_1(W)R_3(W))\right)\\
&&-\frac{1}{24}\left(R_4(w)-E(R_4(W))\right)
+\frac{1}{72}\left(R_3(w)^2-E(R_3(W)^2)\right).
\end{eqnarray*}
\end{proof}

\subsection*{Proof of Theorem \ref{due}}

\begin{proof}
The proof follows the same steps as in Liu {\em et al.} (2014), Section 3, generalized for the use of a SR-posterior (\ref{star}) instead of the classic posterior distribution.

Let $\pi(\theta)$ be a prior distribution for $\theta$ and $p(x|\theta)$ be the conditional distribution of $X$ given $\theta$. Thus, the marginal distribution of $X$ can be written as $p(x)=p(x|\theta)\pi(\theta)/\pi(\theta|x)$, with $\pi(\theta|x)$ the conditional distribution of $\theta$ given $x$.  The functional (\ref{functional}) associated to an $\alpha-$divergence  between the prior $\pi(\theta)$ and the SR-posterior (\ref{star}) can be written as
\begin{eqnarray}
\label{functional2}
T(\pi)&=&
\frac{1}{\alpha(1-\alpha)}\left[ 1-
 \int_{\mathcal{X}}  \left(  \int_{\Theta}   \pi(\theta)^\alpha \pi_{SR}(\theta|x)^{1-\alpha} d\theta \right) p(x) dx 
\right] \nonumber \\
&=&\frac{1}{\alpha(1-\alpha)}\left[ 1- \int_{\Theta} \left( \int_{\mathcal{X}}  \pi(\theta)^\alpha  \pi_{SR}(\theta|x)^{1-\alpha} \pi(\theta|x)^{-1}p(x|\theta)dx \right) \pi(\theta)d\theta
\right] \nonumber \\
\label{func} &=&\frac{1}{\alpha(1-\alpha)}\left[ 1-\int_{\Theta} \pi(\theta)^{\alpha+1}E_\theta\left[\pi_{SR}(\theta|X)^{1-\alpha}\pi(\theta|X)^{-1}\right]d\theta\right]  ,
\end{eqnarray} 
where $E_\theta(\cdot)$ denotes expectation with respect to the conditional distribution of $X$ given $\theta$.
For $\alpha=0$ or $1$, we need to interpret $T(\pi)$ as its limiting value, when it exists.

Now, we apply the shrinkage argument for evaluating the conditional expectation in (\ref{functional2}), $E_\theta\left[\pi_{SR}(\theta|X)^{1-\alpha}\pi(\theta|X)^{-1}\right]$. See Datta and Mukerjee (2004) for details on the shrinkage method. 

Let us first consider the case of $0<|\alpha|<1$.

The first step of the shrinkage method involves fixing a suitable prior distribution $\bar\pi(\theta)$ and calculating the expected value  of $\pi_{SR}(\theta|x)^{1-\alpha}\pi(\theta|x)^{-1}$ with respect to the corresponding posterior $\bar\pi(\theta|x)$, i.e.\ $E^{\bar\pi}\left[\pi_{SR}(\theta|x)^{1-\alpha}\pi(\theta|x)^{-1}|x\right]$. Notice that here we use the classic posterior based on the prior $\bar\pi(\theta)$, i.e. the conditional distribution of $\theta$ given $x$, $\bar\pi(\theta|x)$. Using up to first order the asymptotic expansion (\ref{post-expansion}) for $\pi_{SR}(\theta|x)$ and similar expansions for $\bar\pi(\theta|x)$ and $\pi(\theta|x)$, we obtain
\[
\pi_{SR}(\theta|X)^{1-\alpha}=\left(\frac{n^d|\tilde H|}{(2\pi)^d}\right)^{(1-\alpha)/2} \exp{\left\{ \frac{n(1-\alpha)}{2}(\theta-\tilde\theta)^T\tilde H(\theta-\tilde\theta)\right\}}+O_p(n^{-1/2})
\]
and
\[
\pi(\theta|X)^{-1}\bar\pi(\theta|X)=1
+O_p(n^{-1/2}).
\]
Since $\tilde H$ tends to $G$ as $n\rightarrow \infty$, we have that
\begin{eqnarray*}
E^{\bar\pi}\left[\pi_{SR}(\theta|x)^{1-\alpha}\pi(\theta|x)^{-1}|x\right]
&=&\left(\frac{n^d|\tilde H|}{(2\pi)^d}\right)^{(1-\alpha)/2}
\int_{\Theta} \exp{\left\{ \frac{n(1-\alpha)}{2}(\theta-\tilde\theta)^T\tilde H(\theta-\tilde\theta)\right\}} d\theta\\
&&+O(n^{-1/2})\\
&=&\left(\frac{n^d|G(\theta)|}{(2\pi)^d}\right)^{-\alpha/2}(1-\alpha)^{-d/2}+O(n^{-1/2}).
\end{eqnarray*}

The second step of the shrinkage argument requires to integrate again with respect to the distribution of $X$ given $\theta$ and with respect to the prior $\bar\pi(\theta)$, thus obtaining
\begin{eqnarray*}
\lefteqn{ \int_\Theta E_\theta\left[ E^{\bar\pi}\left[\pi_{SR}(\theta|X)^{1-\alpha}\pi(\theta|X)^{-1}|X\right]\right] \bar\pi(\theta) d\theta }\\
&&=\int_\Theta \left(\frac{n^d|G(\theta)|}{(2\pi)^d}\right)^{-\alpha/2}(1-\alpha)^{-d/2} \, \bar\pi(\theta) \, d\theta +O(n^{-1/2}).
\end{eqnarray*}

Finally, by letting the prior $\bar\pi(\theta)$ go to $\theta$, we obtain
\begin{eqnarray}
\label{expectation}
E_\theta\left[\pi_{SR}(\theta|X)^{1-\alpha}\pi(\theta|X)^{-1}\right]&=&
\left(\frac{n}{2\pi}\right)^{-d\alpha/2}|G(\theta)|^{-\alpha/2}(1-\alpha)^{-d/2}+O(n^{-1/2}).
\end{eqnarray}

For $0<|\alpha|<1$, by substituting (\ref{expectation}) in (\ref{func}) we can see that the selection of a prior $\pi(\theta)$ corresponds to the minimization  with respect to $\pi(\theta)$ of the functional 
\begin{equation}\label{alpha}
\frac{1}{\alpha(1-\alpha)}\int_\Theta \pi(\theta)^{\alpha+1}|G(\theta)|^{-\alpha/2}d\theta.
\end{equation}
Notice that the preceding expression can be interpreted as an increasing transformation of the $(-\alpha)-$divergence between a density that is proportional to  $|G(\theta)|^{1/2}$ and the prior $\pi(\theta)$. Thus, it is minimized if and only if the two densities coincide.

This result  cannot be extended to the case of $\alpha\geq 1$, as is evident from the right-hand-side of (\ref{expectation}). For $\alpha<-1$, (\ref{alpha}) turns out to be equivalent to a decreasing transformation of the $(-\alpha)-$divergence between a density that is proportional to  $|G(\theta)|^{1/2}$ and the prior $\pi(\theta)$. Thus a maximizer for the expected $\alpha-$divergence in this case does not exist.

For $\alpha\rightarrow 0$, following the same steps as for $0<|\alpha|<1$, it can be easily shown that maximization of the average Kullback-Leibler divergence is asymptotically equivalent to minimization of
\[
\int_\Theta \pi(\theta) \log\left(\frac{\pi(\theta)}{|G(\theta)|^{1/2}}\right) d\theta,
\]
which is attained by choosing again a Jeffreys-type prior proportional to $ |G(\theta)|^{1/2}$. Indeed, the preceding expression is equal up to an additive constant to Kullback-Leibler divergence between a density that is proportional to $ |G(\theta)|^{1/2}$ and the prior $\pi(\theta)$.

The case $\alpha=-1$ corresponds to the Chi-square divergence. For this case the proof requires higher order terms in the expansion of the scoring rule posterior distribution and also of both the conditional distributions of $\theta$ given $x$ calculated with respect to the priors $\pi(\theta)$ and $\bar\pi(\theta)$. Thus, we use (\ref{post-expansion}) up to order $O_p(n^{-3/2})$. In particular, 
let $w=n^{1/2}(\theta-\tilde\theta)$ and $w_1=n^{1/2}(\theta-\hat\theta)$, where $\tilde\theta$ is the minimum scoring rule estimator and $\hat\theta$ the maximum likelihood estimator for $\theta$. The different posterior distributions for $w$ and $w_1$ given $x$ can be written as
\begin{eqnarray*}
\pi_{SR} (w|x)=\phi_d(w;\tilde H^{-1})
\left[
1+n^{-1/2}A_1(w) + n^{-1}A_2(w)\right]+O_p(n^{-3/2}),
\end{eqnarray*}
\begin{eqnarray*}
\pi(w_1|x)=\phi_d(w_1;(\hat J^\ell)^{-1})
\left[
1+n^{-1/2}A^\ell_1(w_1) + n^{-1}A^\ell_2(w_1)\right]+O_p(n^{-3/2}),
\end{eqnarray*}
\begin{eqnarray*}
\bar\pi(w_1|x)=\phi_d(w_1;(\hat J^\ell)^{-1})
\left[
1+n^{-1/2}\bar A^\ell_1(w_1) + n^{-1}\bar A^\ell_2(w_1)\right]+O_p(n^{-3/2}),
\end{eqnarray*}
where $A_1$ and $A_2$ are defined as in Theorem \ref{uno}, $J^\ell(\theta)=(-\partial^2\ell/\partial\theta\partial\theta^T)/n$ is the observed information and $A^\ell_1$, $\bar A^\ell_1$, $A^\ell_2$, $\bar A^\ell_2$ are obtained by calculating $A_1$ and $A_2$ with the logarithmic score as scoring rule and $\pi$ and $\bar\pi$ as priors; for instance, 
$$
A^\ell_1(w_1)= \frac{\hat\pi_i}{\hat\pi}w_1^i+\frac{1}{6}\frac{\hat \ell_{ijk}}{n}w_1^iw_1^jw_1^k,
\quad\mbox{and}\quad
\bar A^\ell_1(w_1)= \frac{\hat{\bar\pi}_i}{\hat{\bar\pi}}w_1^i+\frac{1}{6}\frac{\hat \ell_{ijk}}{n}w_1^iw_1^jw_1^k,
$$ 
with $\ell$ the log-likelihood and $\ell_{ijk}$ its third order derivatives. As usual, a tilde or a hat over a quantity indicate that it is calculated at $\tilde\theta$ and $\hat\theta$ respectively.

Following the shrinkage argument, we need first to evaluate 
\begin{eqnarray*}
E^{\bar\pi}\left[\pi_{SR}(\theta|x)^2 \pi(\theta|x)^{-1}|x\right]
&=& n^{d/2} \int \pi_{SR}(w|x)^2  \pi(w|x)^{-1} \bar\pi(w|x) dw . 
\end{eqnarray*}
For doing this, we use the fact that
\begin{eqnarray*}
\pi_{SR} (w|x)^2 &=& (2\pi)^{-d/2} |\tilde H|^{1/2} 2^{-d/2} \phi_d(w;\tilde H^{-1}/2) \\
&&\left[1+n^{-1/2}2A_1(w) + n^{-1}(2A_2(w)+A_1(w)^2)\right]+O_p(n^{-3/2})
\end{eqnarray*}
and
\begin{eqnarray*}
\pi(w_1|x)^{-1} \bar\pi(w_1|x) &=& 
1+n^{-1/2}(\bar A^\ell_1(w_1)-A^\ell_1(w_1)) \\
&&+ n^{-1}(A^\ell_1(w_1)^2+\bar A^\ell_2(w_1)-A^\ell_2(w_1)-\bar A^\ell_1(w_1) A^\ell_1(w_1))  +O_p(n^{-3/2})
\end{eqnarray*}
and we calculate the last expression in  $w_1=w+n^{1/2}(\tilde\theta-\hat\theta)$.

After tedious calculations, we obtain
\begin{eqnarray*}
\lefteqn{E^{\bar\pi}\left[\pi_{SR}(\theta|x)^2 \pi(\theta|x)^{-1}|x\right]=}   \\
&=& (2\pi)^{-d/2}\left(\frac{n}{2}\right)^{d/2} |\tilde H|^{1/2} \left\{1
+ \frac{1}{n^{1/2}} \left(
\frac{\hat{\bar\pi}_{i}}{\hat{\bar\pi}}
-\frac{\hat{\pi}_{i}}{\hat{\pi}}
\right)n^{1/2}(\tilde\theta-\hat\theta)^i \right.\\
&&\left.+ \frac{1}{2n} \left[
\left(-\frac{\hat{\bar\pi}_{ij}}{\hat{\bar\pi}}
+\frac{\hat\pi_{ij}}{\hat\pi}
\right) \hat \bj^{ij}  
\right.\right.\\
&&\left.\left.
+\left( -\frac{\tilde\pi_{ij}}{\tilde\pi}+\frac{\tilde\pi_{i}}{\tilde\pi}\frac{\tilde\pi_{j}}{\tilde\pi}
+2\frac{\tilde\pi_{i}}{\tilde\pi}\frac{\hat{\bar\pi}_{j}}{\hat{\bar\pi}}
-2\frac{\tilde\pi_{i}}{\tilde\pi}\frac{\hat{\pi}_{j}}{\hat{\pi}}
+\frac{\hat\pi_{i}}{\hat\pi}\frac{\hat{\pi}_{j}}{\hat{\pi}}
-\frac{\hat\pi_{i}}{\hat\pi}\frac{\hat{\bar\pi}_{j}}{\hat{\bar\pi}}
+\frac{1}{2}\frac{\hat{\bar\pi}_{ij}}{\hat{\bar\pi}}
-\frac{1}{2}\frac{\hat\pi_{ij}}{\hat\pi}
\right) \tilde h^{ij}  
\right.\right. \\
&& \left.\left.
+\left( \frac{\tilde\pi_{i}}{\tilde\pi}
-\frac{1}{2}\frac{\hat{\bar\pi}_{i}}{\hat{\bar\pi}}
+\frac{1}{2}\frac{\hat{\pi}_{i}}{\hat{\pi}}\right) 
\frac{\tilde S_{khl}}{n} c_{kj}c_{hs}c_{lt}\tilde h^{st}\tilde h^{ij}
+\left(
-\frac{\hat{\bar\pi}_{i}}{\hat{\bar\pi}}
+\frac{\hat{\pi}_{i}}{\hat{\pi}}\right) 
\frac{\tilde \ell_{jhk}}{n} \hat\bj^{hk}\hat\bj^{ij}
\right.\right.\\
&&\left.\left.+ \left(
-\frac{\hat\pi_{ij}}{\hat\pi}+\frac{\hat{\bar\pi}_{ij}}{\hat{\bar\pi}} 
+2\frac{\hat{\pi}_{i}}{\hat{\pi}} \frac{\hat{\pi}_{j}}{\hat{\pi}} 
-2\frac{\hat{\pi}_{i}}{\hat{\pi}} \frac{\hat{\bar\pi}_{j}}{\hat{\bar\pi}} 
\right)
n(\tilde\theta-\hat\theta)^{ij}
\right]+ R \right\}  + O(n^{-1}),
\end{eqnarray*}
where $\hat\bj^{ij}$ are the components of $(\hat J^\ell)^{-1}$ and $R=R(\tilde\theta,\hat\theta)$ is a function that does not involve the prior nor its derivatives.

Now, let $E_\theta(S_{ijk}/n)=B^S_{ijk}(\theta)+o(n^{-1/2})$, $E_\theta(\ell_{ijk}/n)=B^\ell_{ijk}(\theta)+o(n^{-1/2})$ and recall that $E_\theta(\tilde H)=G(\theta)+o(n^{-1/2})$ and $E_\theta(\hat J^\ell)=I(\theta)+o(n^{-1/2})$, where $I$ is the expected information matrix. 
Moreover, note that $E_\theta[(\tilde\theta-\hat\theta)^i ]=o(n^{-1/2})$ and $E_\theta[n(\tilde\theta-\hat\theta)^{ij}]=g^{ij}+\bi^{ij}-2\sigma^{ij}+o(n^{-1/2})$, where $g_{ij}$ and $g^{ij}$ are the components of $G$ and $G^{-1}$ respectively, $\bi^{ij}$ are the components of $I^{-1}$ and $\sigma^{ij}=nCov_\theta(\tilde\theta,\hat\theta)^{ij}$. Furthermore, let $a^S_j=B^S_{khl}c_{kj}c_{hs}c_{lt}g^{st}$, $a^\ell_j=B^\ell_{jhk}\bi^{hk}$.
By  integrating again the above expression with respect to the distribution of $X$ given $\theta$ we get
\begin{eqnarray*}
\lefteqn{E_\theta\left[E^{\bar\pi}\left[\pi_{SR}(\theta|X)^2 \pi(\theta|X)^{-1}|X\right]\right]=}   \\
&=& (2\pi)^{-d/2}\left(\frac{n}{2}\right)^{d/2} |G|^{1/2} \left\{1
+ \frac{1}{2n} \left[
\left(-\frac{\bar\pi_{ij}}{\bar\pi}
+\frac{\pi_{ij}}{\pi}
\right) \bi^{ij}  
\right.\right.\\
&&\left.\left.
+\left( -\frac{3}{2}\frac{\pi_{ij}}{\pi}+\frac{\pi_{i}}{\pi}\frac{\bar\pi_{j}}{\bar\pi}
+\frac{1}{2}\frac{\bar\pi_{ij}}{\bar\pi}
\right) g^{ij}  
\right.\right. \\
&& \left.\left.
+\left(\frac{3}{2}\frac{\pi_{i}}{\pi}
-\frac{1}{2}\frac{\bar\pi_{i}}{\bar\pi}
\right) 
a^S_j g^{ij}
-\left(
\frac{\bar\pi_{i}}{\bar\pi}
-\frac{\pi_{i}}{\pi}\right) 
a^\ell_j \bi^{ij} 
\right.\right.\\
&&\left.\left.+ \left(
-\frac{\pi_{ij}}{\pi}+\frac{{\bar\pi}_{ij}}{{\bar\pi}} 
+2\frac{{\pi}_{i}}{{\pi}} \frac{{\pi}_{j}}{{\pi}} 
-2\frac{{\pi}_{i}}{{\pi}} \frac{{\bar\pi}_{j}}{{\bar\pi}} 
\right)
(g^{ij}+\bi^{ij}-2\sigma^{ij})
\right]+ F \right\}  + O(n^{-1}),
\end{eqnarray*}
where $F=F(\theta)$ is a function that does not involve the prior nor its derivatives.

By integrating again with respect to the prior $\bar\pi(\theta)$ and by letting the prior go to $\theta$, we finally obtain the expected value needed in (\ref{func}) for $\alpha=-1$:
\begin{eqnarray}
\label{expectation-chi}
\lefteqn{E_\theta\left[\pi_{SR}(\theta|X)^2\pi(\theta|X)^{-1}\right]=}\nonumber \\
&=&(2\pi)^{-d/2}\left(\frac{n}{2}\right)^{d/2} |G|^{1/2}\left\{1+\frac{1}{4n}\left[
\left(-3g^{ij} +4\bi^{ij}-4\sigma^{ij}\right)\frac{\pi_{ij}}{\pi}\right.\right.\nonumber\\
&&\left.\left. +\left( 3a^S_jg^{ij}+2a^\ell_j\bi^{ij}+|G|^{-1}\frac{\partial |G|}{\partial\theta^j}(g^{ij}+2\bi^{ij}-4\sigma^{ij})+2 \frac{\partial (g^{ij}+2\bi^{ij}-4\sigma^{ij})}{\partial\theta^j} \right)\frac{\pi_i}{\pi}
\right.\right.\nonumber\\
&&\left.\left. +2g^{ij}\frac{\pi_{i}}{\pi}\frac{\pi_{j}}{\pi} +M \right]\right\}+O(n^{-1}),
\end{eqnarray} 
where  $M=M(\theta)$ is a function that does not involve the prior nor its derivatives.

Substituting (\ref{expectation-chi}) in (\ref{func}) with $\alpha=-1$, we find that  maximization of the average Chi-square divergence is equivalent to maximization of 
\begin{eqnarray*}
\lefteqn{\int_\Theta |G|^{1/2} \left[\left(-3g^{ij} +4\bi^{ij}-4\sigma^{ij}\right)\frac{\pi_{ij}}{\pi} +2g^{ij}\frac{\pi_{i}}{\pi}\frac{\pi_{j}}{\pi}
 \right.}\\
 &&
\left. +\left(3a^S_jg^{ij}+2a^\ell_j\bi^{ij}+|G|^{-1}\frac{\partial |G|}{\partial\theta^j} (g^{ij}+2\bi^{ij}-4\sigma^{ij} )
+2 \frac{\partial (g^{ij}+2\bi^{ij}-4\sigma^{ij})}{\partial\theta^j} \right)\frac{\pi_i}{\pi}
 \right] d\theta.
\end{eqnarray*}

Let $y=y(\theta)=(y_1(\theta),\dots,y_{d}(\theta))^T$, with $y_i(\theta)=\pi_i(\theta)/\pi(\theta)$, $i=1,\dots, {d}$, and let $y'$ be the matrix of partial derivatives of $y$ with respect to $\theta$, i.e.\ $y'=(y_{ij})_{ij}$, $y_{ij}=\partial y_i/\partial\theta^j$, $i,j=1,\dots { d}$. Then, $\pi_{ij}/\pi=y_{ij}+y_iy_j$ and the quantity to be maximized can be rewritten as
\begin{eqnarray*}
\lefteqn{\int_\Theta |G|^{1/2} \left[\frac{}{} \left(-3g^{ij} +4\bi^{ij}-4\sigma^{ij}\right)y_{ij}+\left(-g^{ij} +4\bi^{ij}-4\sigma^{ij}\right)y_iy_j\right.}\\
 &&\left.+\left(3a^S_jg^{ij}+2a^\ell_j\bi^{ij}+|G|^{-1}\frac{\partial |G|}{\partial\theta^j}(g^{ij}+2\bi^{ij}-4\sigma^{ij} )
 +2 \frac{\partial (g^{ij}+2\bi^{ij}-4\sigma^{ij})}{\partial\theta^j} \right)y_i 
 \right] d\theta.
\end{eqnarray*}
Let us denote the integrand function by $U(\theta,y,y')$. The solution to the maximization problem is found by solving the system of Euler-Lagrange equations: 
\begin{equation*}
\label{EL}
\frac{\partial U}{\partial y_i}-\sum_{j=1}^d\frac{\partial}{\partial\theta^j}\left( \frac{\partial U}{\partial y_{ij}} \right)=0, \quad\quad i=1,\dots,{d}.
\end{equation*}
After some calculations, we obtain the solution to the variational problem as
\begin{eqnarray*}
y(\theta)=\frac{1}{4}\left[
6a^S G^{-1} + 4a^\ell I^{-1}+ |G|^{-1}\frac{\partial |G|}{\partial\theta}(5G^{-1}-4\Sigma)
+2G \nabla\cdot (5G^{-1}-4\Sigma) \right] \Gamma,
\end{eqnarray*}
where $a^S=(a^S_1,\dots,a^S_{d})^T$, $a^\ell=(a^\ell_1,\dots,a^\ell_{d})^T$, $\Sigma=(\sigma^{ij})_{ij}$, $\Gamma=(G^{-1}-4I^{-1}+4\Sigma)^{-1}$, $\nabla\cdot G^{-1}=
(\partial g^{1j}/\partial\theta^j, \dots, \partial g^{{d} j}/\partial\theta^j)^T$ and $\nabla\cdot \Sigma=
(\partial \sigma^{1j}/\partial\theta^j, \dots, \partial \sigma^{{d} j}/\partial\theta^j)^T$.
\end{proof}


\section*{References}

\begin{description}
\item
Barndorff-Nielsen, O.E. (1976). Plausibility inference. {\em J.\ Roy.\ Statist.\ Soc.} B, {\bf 38}, 103--131.
\item
Basu, A., Harris, I.R., Hjort, N.L., Jones, M.C. (1998). Robust and efficient estimation by minimising a density power divergence. {\em Biometrika}, {\bf 85}, 549--559.
\item
Berger, J.O. (2006). The case for objective Bayesian analysis. {\em Bayesian Analysis}, {\bf 1}, 1--17.
\item
Berger, J.O.,  Bernardo, J.M. (1992). On the development of reference priors (with discussion). In {\em Bayesian Statistics} 4 (J.M.\ Bernardo, J.O.\ Berger, A.P.\ Dawid and A.F.M.\ Smith, eds.), 35--60. Oxford Univ.\ Press.
\item
Berger, J.O., Bernardo, J.M., Sun, D. (2009). The formal definition of reference priors. {\em Journal of the American Statistical Association}, {\bf 107}, 636--648.
\item
Berger, J.O., Bernardo, J.M., Sun, D. (2012). Overall objective priors. {\em Bayesian Analysis}, {\bf 10}, 189--221.
\item 
Bernardo, J.M. (1979). Reference posterior distributions for Bayesian inference (with discussion). {\em J.\  Roy.\ Statist.\ Soc.} B, {\bf 41}, 113--147.
\item
Bernardo, J.M. (2005). Reference analysis. In: {\em Handbook of Bayesian Statistics}, {\bf 25} (D.K. Dey and C.R. Rao, eds.), North-Holland, Amsterdam: Elsevier, 17--90.
\item
Brier, G.W. (1950). Verification of forecasts expressed in terms of probability. {\em Mon. Weather Rev.}, {\bf 78}, 1--3.
\item
Chandler, R.E., Bate, S. (2007). Inference for clustered data using the independence loglikelihood. {\em Biometrika},  {\bf 94}, 167--183.
\item
Chang, I., Mukerjee, R. (2008). Bayesian and frequentist confidence intervals arising from empirical-type likelihoods. {\em Biometrika}, {\bf 95}, 139--147.
\item
Corcuera, J.M., Giummol\`e, F. (1998). A characterization of monotone and regular divergences. {\em Ann.\ Inst.\ Statist.\ Math.}, {\bf 50}, 433--450.
\item
Datta, G.S., Mukerjee R. (2004). {\em Probability Matching Priors: Higher Order Asymptotics}. Lecture Notes in Statistics 178. Springer.
\item
Dawid, A.P. (1986). Probability forecasting. In: {\em Encyclopedia of Statistical Sciences} (S. Kotz, N. L. Johnson, and C. B. Read eds.), 210--218. Wiley-Interscience.
\item
Dawid, A.P. (2007). The geometry of proper scoring rules. {\em Ann.\ Inst.\ Statist.\ Math.}, {\bf 59}, 77--93.
\item
Dawid, A.P., Lauritzen, S.L. (2005). The geometry of decision theory. In: {\em Proceedings of the Second International Symposium on Information Geometry and its Applications}, 22--28. University of Tokyo.
\item
Dawid, A.P., Musio, M. (2014). Theory and Applications of Proper Scoring Rules. {\em Metron}, {\bf 72}, 169--183.
\item
Dawid, A.P., Musio, M. (2015).  Bayesian model selection based on proper scoring rules (with discussion). {\em Bayesian Analysis}, {\bf 10}, 479--521. 
\item
Dawid, A.P., Musio, M., Ventura, L. (2016). Minimum scoring rule inference. {\em Scand.\ J.\ 
Statist.}, {\bf 43}, 123--138.
\item
Farcomeni, A., Ventura, L. (2012). An overview of robust methods in medical research. {\em Statistical Methods in Medical Research}, {\bf 21}, 111--133.
\item
Godambe, V.P. (1960). An optimum property of regular maximum likelihood estimation. {\em Ann.\
Math.\ Statist.}, {\bf 31}, 1208--1211.
\item
Good, I.J. (1952). Rational decisions. {\em J.\  Roy.\ Statist.\ Soc.} B, {\bf 14}, 107--114.
\item
Ghosh, M. (2011). Objective Priors: An Introduction for Frequentists. {\em Stat.\ Sci.}, {\bf 26}, 187--202.
\item
Ghosh, M., Basu, A. (2013). Robust estimation for independent non-homogeneous observations using density power divergence with applications to linear regression. {\em Electr.\ J.\ Statist.}, {\bf 7}, 2420--2456.
\item
Ghosh, M., Basu, A. (2016). Robust Bayes estimation using the density power divergence. {\em Ann.\ Inst.\ Stat.\ Math.}, {\bf 68}, 413--437.
\item
Ghosh, M., Mergel, V., Liu, R. (2011). A general divergence criterion for prior selection  {\em Ann. Inst. Stat. Math.} 43--58. 
\item
Ghosh, J.K., Mukerjee, R. (1991). Characterization of priors under which Bayesian and frequentist Bartlett corrections are equivalent in the multi-parameter case. {\em J.\ Mult.\ Anal.}, {\bf 38}, 385--393.
\item
Greco, L., Racugno, W., Ventura L. (2008). Robust likelihood functions in Bayesian inference. {\em J.\ Statist.\ Plann.\ Inf.}, {\bf 138}, 1258--1270.
\item
Hampel, F.R., Ronchetti, E.M., Rousseeuw, P.J., Stahel, W.A. (1986). {\em Robust Statistics: The Approach Based on Influence Functions}. Wiley, New York.
\item
Heritier, S., Cantoni, E., Copt, S., Victoria-Feser, M.P. (2009). {\em Robust methods in biostatistics}. Wiley, Chichester, UK.
\item
Huber, P.J., Ronchetti, E.M. (2009). {\em Robust Statistics}. Wiley, New York.
\item
Hyv\"arinen, A. (2005). Estimation of non-normalized statistical models by score matching. {\em Journal of Machine Learning Research}, {\bf 6}, 695--709.
\item
Hyv\"arinen, A. (2007). Some extensions of score matching. {\em Computational Statistics and Data Analysis}, {\bf 51}, 2499--2512. 
\item
Jeffreys, H. (1961). {\em Theory of Probability}. Oxford University Press, New York.
\item
Lazar, N.A. (2003). Bayes empirical likelihood. {\em Biometrika}, {\bf 90}, 319--326.
\item
Leisen, F., Villa, C., Walker, S.G. (2017). On a global objective prior from score rules. arXiv: 1706.00599v1
\item
Lin, L. (2006). Quasi Bayesian likelihood. {\em Statistical Methodology}, {\bf 3}, 444--455.
\item
Liu, R., Chakrabarti, A., Samanta, T., Ghosh, J.K., Ghosh, M. (2014). On divergence measures leading to Jeffreys and other reference priors. {\em Bayesian Analysis}, {\bf 9}, 331--370.
\item
Machete, R. (2013). Contrasting probabilistic scoring rules. {\em J.\ Statist.\ Plann.\ Inf.}, {\bf 143}, 1781--1790.
\item
Mameli, V., Musio, M., Ventura, L. (2017). Bootstrap adjustments of signed scoring rule root statistics. {\em Comm.\ Statist.\ - Simul.\ Comput.}, to appear.
\item
Mameli, V., Ventura, L. (2015). Higher-order asymptotics for scoring rules. {\em J.\ Statist.\ Plann.\
 Inf.}, {\bf 165}, 13--26.
\item
Mardia, K.V., Jupp, P.E. (2000). Directional statistics. Wiley, London. 
\item
Mardia, K.V., Kent, J.T., Laha, A.K. (2016). Score matching estimators for directional distributions. ArXiv:1604.08470v1.
\item
Musio, M., Mameli, V., Ruli, E., Ventura, L. (2017). Bayesian Inference for directional data through ABC and homogeneous proper scoring rules. {\em Proceeding of 61$st$ ISI World Statistics Congress}, to appear.
\item
Pace, L., Salvan, A., Sartori, N. (2011). Adjusting composite likelihood ratio statistics. {\em Statist.\ Sin.}, {\bf 21}, 129--148.
\item
Parry, M., Dawid, A.P., Lauritzen, S.L. (2012). Proper local scoring rules. {\em Ann.\ Statist.}, {\bf 40}, 561--592.
\item
Pauli, F., Racugno, W., Ventura, L. (2011). Bayesian composite marginal likelihoods. {\em Statist.\ Sin.}, {\bf 21}, 149--164.
\item
Riani, M., Atkinson, A.C., Perrotta, D. (2014). A parametric framework for the comparison of methods of very robust regression. {\em Statistical Science}, {\bf 29}, 128--143.
\item
Ribatet, M., Cooley, D. and Davison, A.C. (2012). Bayesian inference from composite likelihoods, with an application to spatial extremes. {\em Statist.\ Sin.}, {\bf 22}, 813--845.
\item
Ruli, E., Sartori, N., Ventura, L. (2016). Approximate Bayesian Computation with composite score functions. {\em Statist.\ Comp.}, {\bf 26}, 679--692.
\item
Schennach, S. (2005). Bayesian exponentially tilted empirical likelihood. {\em The Annals of Statistics}, {\bf 35}, 634--672.
\item
Smith, E.L., Stephenson, A.G. (2009). An extended Gaussian max-stable process model for spatial extremes. {\em J.\ Statist.\ Plann.\ Inf.}, {\bf 139}, 1266--1275.
\item
Tsallis, C. (1988). Possible generalization of Boltzmann-Gibbs statistics. {\em J.\ Statist.\ Physics}, {\bf 52}, 479--487.
\item
Varin, C., Reid, N. and Firth, D. (2011). An overview of composite likelihood methods. {\em Statist.\ Sin.}, {\bf  21}, 5--42.
\item
Ventura, L., Cabras, S., Racugno, W. (2010). Default prior distributions from quasi- and quasi-profile likelihoods. {\em J.\ Statist.\ Plann.\ Inf.}, {\bf 140}, 2937--2942.
\item
Ventura, L., Racugno, W. (2016). Pseudo-likelihoods for Bayesian inference. In: {\em Topics on Methodological and Applied Statistical Inference, Series Studies in Theoretical and Applied Statistics}, Springer-Verlag, 205--220.
\item
Walker, S.G. (2016). Bayesian information in an experiment and the Fisher information distance. {\em Statistics and Probability Letters}, {\bf 112}, 5-9.
\item
Yang, Y., He, X. (2012). Bayesian empirical likelihood for quantile regression. {\em The Annals of Statistics}, {\bf 40}, 1102--1131.
\end{description}

\end{document}